\documentclass[12pt,preprint]{aastex}  
\usepackage{graphics}

\slugcomment{Version of \today}
\shortauthors{Smith}
\shorttitle{Far-UV atlas of B stars} 
\begin{document} 

\newcommand{\iue}{{\it IUE}}
\newcommand{\stis}{{\it STIS}}
\newcommand{\fuse}{{\it FUSE}}
\newcommand{\cope}{{\it Copernicus}}
\newcommand{\dl}{$\lambda$}
\newcommand{\no}{\noindent}

\title{A Detailed Far-Ultraviolet Spectral Atlas of Main Sequence B Stars }

\author{Myron A. Smith,} 
\affil{Catholic University of America,\\
        3700 San Martin Dr.,
        Baltimore, MD 21218;           \\
        msmith@stsci.edu,}

\clearpage

 \begin{abstract}

  We have constructed a {\em detailed} spectral atlas covering the wavelength 
region 930\,\AA~ to 1225\,\AA~ for 10 sharp-lined B0-B9 stars near the
main sequence.  Most of the spectra we assembled are from the archives of the
{\it FUSE} satellite, but for nine stars wavelength coverage above 1188\,\AA~ 
was taken from high-resolution {\it IUE} or echelle {\it HST/STIS} spectra.
To represent the tenth star at  type B0.2\,V we
used the {\it Copernicus} atlas of $\tau$\,Sco. 
We made extensive line identifications in the region 949\,\AA~ to 1225\,\AA~
of all atomic features having published oscillator strengths at types B0,
B2, and B8. These are provided as a supplementary data product, 
-  hence the term {\em detailed} atlas. Our list of found 
features totals 2288, 1612, and 2469 lines, respectively. We were
able to identify 92\%, 98\%, and 98\% of these features with known
atomic transitions with varying degrees of certainty in these spectra. 
The remaining lines do not have published oscillator strengths. 
Photospheric lines account for 94\%, 
87\%, and 91\%, respectively, of all our identifications, with the 
remainder being due to interstellar (usually molecular H$_2$) lines.  
We also discuss the numbers of lines with respect to the distributions 
of various ions for these three most studied spectral subtypes. 
A table is also given of 167 least blended lines that 
can be used as possible diagnostics of physical conditions in B star 
atmospheres.
\end{abstract}
 

\clearpage

\section{Introduction}
\label{int}

  To understand the physical conditions of O and B stars and their
immediate environments there can be no better mode of study than  
high-dispersion ultraviolet spectroscopy. The spectral energy 
distributions of hot stars peak in this regime, and it is also here that 
absorption lines arising from a wide range of excited atomic states 
appear in high concentration.  These lines provide a rich and rather 
unexplored area for the study of markers of atmospheric 
temperature, pressure, kinematics, and chemical composition. 
In addition, atoms undergoing far-UV transitions ``see" down to deep
layers of the atmosphere because of the local relative transparency. 
The resulting intense UV radiation field is responsible for
creating departures from LTE in these transitions and influencing
the calibration of spectral diagnostics. Once atmospheres of single 
hot stars are understood, this knowledge can be extended to the 
interpretation of spectra of distant ensembles of hot stars and star 
formation regions.  

  At about the turn of this century the era dawned in which specific 
far-UV spectral lines observed in OB clusters embedded in highly 
redshifted galaxies can be studied from the ground at optical 
wavelengths (e.g., de Mello 2000 et al., Pettini et al.  2001, Dawson
et al. 2002, Pettini et al. 2002, Mehlert et al. 2002, Croft et al.
2002, Robert et al. 2003, Shapley et al. 2003, de Mello et al. 2004).
In the two decades leading up to and including this milestone, enormous 
progress was made in understanding the atmospheres and origin of OB stars. 
First came high-resolution spectroscopic observations of the middle-UV 
during the extended period (1978-1996) of the 
{\it International Ultraviolet Explorer (IUE)} satellite. Next, the 
1999 launch of the {\it Far Ultraviolet Spectroscopic Explorer (FUSE),} 
brought access to the far-UV for hundreds of hot stars in the Milky Way 
and Magellanic Clouds.  This era of grand exploration and surveys 
closed with the decommissioning of {\it FUSE} in 2007.  
The final reprocessing of the {\it FUSE} spectra (covering the wavelength
region 920\,\AA~ to 1188\,\AA) now permits the examination of a wealth of 
homogeneous high-quality material.

  For O and B stars one consolidation of these gains was 
the addition of a fine set of middle-UV and far-UV spectral atlases. 
Examples of the former are the {\it Copernicus} atlas of $\tau$\,Sco 
(Rogerson et al. 1977) and {\it IUE} pictorial atlases of O and B stars 
(Walborn, Nichols-Bohlin, \& Panek 1985, Rountree \& Sonneborn 1993, 
Walborn, Parker, \& Nichols 1995).
Pellerin et al. (2002) inaugurated a second group of digitized atlases
with their atlases of O and early-B type stars of all luminosity classes. 
This work had the goal of exhibiting the general
behavior of the strong photospheric lines and the so called UV 
wind lines with effective temperature T$_{eff}$ and luminosity class.
Valuable supplemental information about the molecular H$_2$ lines formed
in the interstellar medium (ISM) was also provided. From the point
of view of stellar researchers, these features contaminate many of the
far-UV photospheric lines of most Galactic stars. This atlas was closely 
followed by a second spectral atlas on OB stars in the two Magellanic 
Clouds by the same group (Walborn et al. 2002). This work concentrated 
on the behavior of wind lines with respect to metallicity as well as 
temperature and luminosity.  In addition to these atlases, Blair et al. 
(2009) have published a compendium of {\it FUSE} spectra of hot stars 
in the Magellanic Clouds. This atlas focused on the identification 
of far-UV resonance lines arising from the ISM that are sometimes
found at multiple velocities. Specialized atlases are now underway
for the purposes of displaying the behavior of specific elements in
particular types of stars. For example, an atlas depicting lines of
heavy metal elements in {\it IUE} spectra of late-B and early B stars
has recently been published by Adelman et al. (2004). 

  The first far-UV high-dispersion spectral coverage published for a 
B star was the {\it Copernicus} atlas of $\tau$\,Scorpii (B0.2\,V; 
Walborn 1971) by Rogerson \& Upson (1977; ``RU77"). 
Rogerson \& Ewell (1985; ``RE85") published a detailed tabulation
of photospheric and ISM lines identified in this atlas.
The RE85 work was undertaken at
a time when spectral synthesis tools were not commonly available, 
and when reliable identifications lay in the hands of a comparative 
handful of experienced spectroscopists who were 
also specialists in atomic physics. Even so, in the absence of 
commonly available spectral syntheses programs at that time, 
it was difficult to make wholesale 
line identifications without significant numbers of errors. This 
situation has changed dramatically in the intervening years with
the development of spectral line synthesis tools that make use
of extensive atomic line libraries (as well as the organization
of the line libraries themselves).

  Inspired by the {\it Copernicus} atlas, we decided to
address the need for the identification of lines in
high-dispersion far-UV spectra along the B main sequence. 
Thus, the first goal of our work is to select and provide
far-UV spectral data for main sequence B0 to B9 stars.
The atlas spectra should be sharp-lined enough 
to permit the resolution of a majority of dominant individual
atmospheric lines and to provide spectral data products in a 
common, continuous-spectrum format ranging in wavelength from near the 
short-wavelength limit of the {\it FUSE} regime to the red wing (1225\,\AA)
of Lyman\,$\alpha$.  Such an assemblage requires 
not only the cojoining of all {\it FUSE} spectral detector
segments but also (for B0) of second-order scans of {\it Copernicus's}
``U1" multiplier, and echelle spectra 
from {\it IUE} or the {\it Hubble Space Telescope Imaging Spectrograph 
(STIS)}. This is necessary to extend our coverage from the {\it FUSE} 
long-wavelength limit of 1188\,\AA~ to 1225\,\AA, as justified below.

Our second goal is to choose three spectra among our atlas sample 
that capture the low, high, and 
midpoints of the excitation range of ions in B star atmospheres 
and to identify all possible absorption lines that make a noticeable
contribution to these spectra according to published atomic
line libraries.
This addition changes the
emphasis of the atlas from the classical high-level montage
presentation to a ``detailed" attribution of the spectral features.
Our third goal is to discuss those metallic lines identified from
the second goal in the context of diagnostics of effective 
temperature, luminosity, and heavy element abundances.

This paper is organized as follows. The selection of spectra and
the methodology for data conditioning and line identification
are set forth in $\S$2. Section $\S$3 displays portions of the 
atlas and a detailed list of all (several thousand) of our line 
identifications as well as a list of ``clean" lines across 
much of the B-star domain.
We also discuss an example of the important C\,III 1176\,\AA~ 
region used as a temperature diagnostic for this spectral type.
In $\S$4 we give relevant statistics from our identifications 
for the B0, B2, and B8 exemplars we call spectral templates. 
We provide a comparison of our results with an earlier critique 
of the Rogerson \& Ewell (1985) line attributions for the $\tau$\,Sco 
atlas. We finally give commentary on a number of possible spectral 
markers for physical conditions in these stars' atmospheres.

\section{Preparation of the Atlas}
\label{obsn}

\subsection{Spectral coverage} 

  The primary purpose of this atlas is to provide a continuous spectrum 
from 930\,\AA~ to 1225\,\AA~ of every spectral subtype of a Galactic 
B-type star near the main sequence. This means that 
the atlas is based largely but not exclusively on {\it FUSE} spectra, a
requirement introduces some complications because it necessitates that 
data from other instruments be used for coverage beyond the long-wavelength 
limit of {\it FUSE} at 1188\,\AA. This in turn imposes changes at this
wavelength in instrumental sampling, resolution, and, most important of 
all, a shift to a spectrum of a second star that often reflects slightly 
different physical properties. Our rationale for extending the atlas 
wavelength coverage is, first, that we wanted to bridge the broad ``middle 
UV" and ``far-UV" regimes empirically defined by the coverages of {\it IUE} 
and {\it FUSE} ultraviolet explorer missions. One may define this bridge
region in practice as the interval between the long-wavelength limit of 
{\it FUSE} coverage and 1225-6\,\AA,~ where the roll-off in sensitivity 
at short wavelengths commences for the {\it IUE} Short Wavelength (SWP) 
camera.  Second, our aim is to include the Lyman $\alpha$ line, which 
along with higher Lyman members covered by the {\it FUSE} detectors, 
is an important ingredient in determining the radiative and thermal 
equilibria of a star's atmosphere and thus its physical structure.
Note that the innermost 6-8 \AA ngstroms in the Lyman $\alpha$
lines are dominated by an ISM component in spectra of B stars
located in the Galactic plane (e.g., Bohlin 1975). Thus by extending
atlas coverage to 1225\,\AA~ we can ensure coverage of photospheric lines 
on both sides of the  Lyman $\alpha$ line core.  Important spectral 
diagnostics of temperature and ionization are also included in this 
region, including sometimes strong resonance lines of Si\,III 
(1206\,\AA) and Si\,II (1193-1194\,\AA).  Finally, the identification 
of metallic lines in this region may help distinguish spectral lines
formed in atmospheres in stars embedded in redshifted galaxies from 
those of the local hydrogenic Lyman forest.

\subsection{Atlas star selection} 
\label{opt}

The selection of atlas stars required that the spectra have high
signal-to-noise ratios and exhibit sharp and single lines. Whenever possible, 
they must also represent normal (solar-like) photospheric abundances and 
not suffer much ISM absorption (in practice this means E$(B-V)$ $<$ 0.4).
To find suitable far-UV spectra we combed through all B-type III-V class
material in the MAST/{\it FUSE} archive\footnote{The Multi-Mission 
Archive at Space Telescope Science Institute (STScI). 
The STScI is operated by the Association of Universities for Research in 
Astronomy, Inc., under NASA contract NAS5-26555. Support for non-HST MAST 
data is provided by NASA Office of Space Science via grant NAS5-7584.} 
and identified stars with sharp-lined spectra. 
For coverage above the long-wavelength limit of {\it FUSE} 
spectra, we included spectra from either the {\it IUE} or 
HST/{\it STIS} archive. To follow the discussion on specific stars below,
the reader is referred to the final list of our selected atlas stars 
in Table\,1.

As a unique case, and being sharp-lined and bright, 
the star $\tau$\,Scorpii is a ``ready made" representative of type B0\,V. 
(Note, however, that its wavelength coverage starts at 948.7, and not 
930, \AA ngstroms. We also note that wavelength corrections have been
introduced following an Erratum in the original Rogerson \& Upson paper.)
For other stars we inspected spectral previews of several hundred 
stars and in some instances the FITS data themselves to 
make nine additional selections. Spectral types were taken from the 
Skiff (2007) catalog, which in the case of multiple assignments
fortunately showed at most a small range in range of estimated 
spectral subtypes.  Our star selections required a few compromises.  
Notably, both of our latest type stars, HD\,182308 
and HD\,62714, have peculiar abundances. HD\,182308 has 
been variously classified as B8\,V (Floquet 1970) and B9p(Hg)Mn (Cowley
1968). A detailed inspection of its far-UV line strengths confirms
that this star's atmosphere has anomalous abundances consistent with 
the HgMn class class of peculiar B stars.
HD\,62714 has been classified as B9/B9.5HeB7Vp: (Houk \&
Cowley 1975). Because we noticed that some far-UV lines of chromium or
manganese appeared anomalous in strength, the classifications of these
two stars as chemically peculiar seems appropriate.  
For purposes of this atlas, we designated 
HD\,182308 as B8\,Vp and HD\,62714 as B9\,Vp.
With these exceptions to our stipulated ideal criteria, we list 
the star selections for this far-UV spectral atlas in Table\,1. 

  The ordering of stars in this table was facilitated by the 
progression with spectral subtype of the important C\,III 1176\,\AA~ 
complex (see discussion in $\S$\ref{node1176}), strengths of a few
Fe-line (principally complexes), as well as
the effective temperatures reported by Kilian (1994; for
$\tau$\,Sco), Glagolevskij (1994), by T$_{eff}$ values determined
by Fitzpatrick \& Massa (1999) and Morel \& Butler (2006),
and by the $uvbyH\beta$ 
temperature calibration of B stars of Napiwotzki, 
Schoenbrunner \& Wenske (1993).
Our selections left an
unfilled gap in our spectral subtypes at B7. However, this 
gap in T$_{eff}$ is rather small as the temperatures of the
HD\,182308 and HD\,62714 both seem to be slightly high for 
stars having these spectral subtypes. Reddening (E(B-V)) values, 
which generally correlate with ISM column density, are taken from
several sources in the literature. Values for the bright stars
we used for ``adjunct" ($>$1188\,\AA) spectra, are $\le$ 0.10. 

  Panchromatic data from the same instrument ({\it Copernicus}) 
were available for only one star, $\tau$\,Sco. For the other 
spectral types we used observations for the same star both below
and above 1188\,\AA~ wherever possible. However, as noted in $\S$2.1, we 
were obliged to provide this extension by using coadded echelle {\it IUE} 
spectra for another star of virtually the same spectral type in 
the following cases: for type B1 ($\xi^{1}$\,CMa), B2 
($\gamma$\,Peg), B2-B3 ($\zeta$\,Cas), B3-B4 ($\iota$\,Her), 
and B8p-B8 ($\xi$\,Oct.) For B5 (HD\,94144),  B6\,V (HD\,30122), 
and (B9p) HD\,62714 echelle {\it STIS} spectra were utilized.  
Because the temperature of our B1 star HD\,113012 is already nearly as 
high as $\tau$\,Sco (and to exclude it from the atlas would risk instead 
adopting another spectrum with much broader lines), it was expedient 
to use {\it Copernicus} of $\tau$\,Sco a second time to extend the
coverage of HD\,113012 for the limited range 1188\,\AA~ to 1225\,\AA.\footnote{The
MAST {\it IUE} archive contains spectra of another B0.5\,III star, 1\,Cas, 
This star is a slowly rotating near-twin of HD\,102475, which was the star for
which we chose {\it FUSE} data for this spectral type.
With only three SWP camera spectra available, the coadded spectrum of 1\,Cas 
contains considerable noise, and we rejected it for this reason. 
We mention this in case interested readers prefer to consult an alternative
long wavelength spectrum to $\tau$\,Sco for type B\,0.5.}
The details of the adjunct spectra are noted as parenthesized, second-line
entries in Table\,1.

\subsection{Data properties and handling} 

\subsubsection{Basic data properties}

``Final" data reprocessings of {\it IUE} and {\it STIS} echelle 
spectra were completed in 1997 and 2006, respectively.
(The recent refurbishment of the HST by the Space Shuttle Atlantis 
ensure a new generation of STIS data will follow.)
The {\it FUSE} data were reprocessed with CalFUSE version 3.2 
(Dixon et al. 2007) during 2007-8.  These spectra were ingested
into the MAST archive,
and we retrieved them from this facility.


  The {\it FUSE} spacecraft utlized four independent telescopes and 
spectrographs. Paraphrasing the {\it FUSE Archival Instrument Handbook}
(Kaiser \& Kruk 2009),
each of the telescopes illuminated its own holographic 
diffraction grating/camera mounted on a Rowland circle spectrograph  
and fed light to one of two far-UV microchannel plate detectors that
illuminated two microchannel plate detectors via LiF and SiC coated mirrors.  
Each of the detectors recorded spectra from a pair of these optical channels,
one each from focused a camera mirror coated with LiF or SiC and therefore 
optimized for a limited wavelength range.
Nearly complete far-UV coverage of the spectrum is provided by
two nearly identical ``sides" of the instrument, each of which includes
two pairs of LiF and SiC detectors.  Flux at almost all wavelengths
is recorded by at least two segments
(the region 1015-1075\,\AA~ is covered in four segments). 
Table\,2 lists in italics 
the coverages of each of the {\it FUSE} detector segments. 

Both {\it IUE} and {\it HST/STIS} data were obtained through large science
apertures with an echelle grating optimized for high orders. References
for this instrument and data processing are given, respectively, in
Garhart et al. (1997) and Kim Quijano et al. (2007).

  The spectral resolutions (full width half maxima) 
of these spectra, as taken from RU77 or the cited data handbooks 
are as follows: {\it Copernicus} 12-15 km\,s$^{-1}$, 
{\it FUSE} 15 km\,s$^{-1}$, {\it STIS} 12-15 km\,s$^{-1}$, 
and {\it IUE} 30 km\,s$^{-1}$. Thus, the resolution of these instruments 
is approximately matched to the rotational broadening criterion we 
imposed in our star selection.

\subsubsection{Conditioning of {\it FUSE} spectra}

  The conditioning steps for {\it FUSE} data were comparatively
more complex, first, because of the well known nonlinear excursions
of the wavelength scale. These effects 
are largely due to electron repulsion in the detector that can distort
the faithful deposition of photoelectrons. Such excursions
are typically not robust with time and cannot be modeled reliably.  
In addition to this, systematic shifts due to the positioning of the star 
image in a large science aperture are present. 
A third problem is the presence of optical vignetting (``worms") 
across the detector field that caused artificial and uncharacterizable 
depressions in certain detector segments, particularly 
the LiF 1B segment (see Chapter~4 of the FUSE Instrument Handbook; 
Kaiser \& Kruk 2009). Because several instrumental idiosyncracies appear in
{\it FUSE} detector Sides 1 and 2 spectra, most researchers have learned 
to work with the segments of these two sides separately and compare them 
in the end as if they were independent observations. We will do the same 
here, as we describe the steps to our building two cojoined spectra from 
the 4 detector segment pairs.

   In almost all cases multiple exposures had been taken of our target
stars. Therefore, for each of the eight detector segments our first 
conditioning steps were to cross-correlate to the nearest pixel, 
co-weight, and add the individual spectral exposures.  
Coweighting was implemented by means 
of a simple pixel-to-pixel fluctuation metric used to compute signal-to-noise 
ratios (Stoehr 2007). In this coweighting step we omitted any 
spectra (e.g., resulting from incorrect pointings) with weights less 
than ${\frac 13}$ the of the maximum weighted observation.  
Visual inspection then verified that all spectra had been shifted 
to the nearest pixel of the fiducial (first) observation in the series.

   Except for linear interpolation over subpixel scales noted below, 
our only modifications to the fluxes were to substantially remove the
flux depression due to the worm in the region from 1130\,\AA~ to 1160\,\AA~ 
of the LiF 1B segment. This was done by passing a high order filter having
the same degree over the 1B and 2A segment fluxes and then interpolating 
to the original pixel scale. The 1B worm was ``removed" by dividing 
of the 2A spectrum polynomial fit by the 1B fit in the wavelength region 
affected by the worm and applying the quotient to the original 1B spectrum.

   It was necessary to devote considerable attention to our correction of 
the wavelength scale. The raw wavelengths exhibits frequent departures of
0.03\,\AA~ or larger from linearity over occasionally even a few \AA ngstroms. 
This was accomplished by first determining a trial linear wavelength 
calibration (generally, by a small, and always subpixel, shift of the 
spectrum) to match computed Lyman and Werner transitions from a single 
rotational state of the zeroth vibrational level to the ground state of 
molecular H$_2$ lines in the interstellar medium (ISM). The parameters 
of these transitions were provided by an on-line tool created as part
of the ``H2ools" project (McCandliss 2003). 
This tool allowed us to identify these molecular ISM features in 
all our spectra and and to coplot them on a common wavelength scale.
Our procedures permitted us to detect wavelength scale nonlinearities 
and correct them by imposing shifts on a subpixel scale (0.01\,\AA) 
over adjacent wavelength segments to bring the observations into 
conformity with the theoretical H$_2$ line ``comb." This procedure 
was repeated for all segments of the two {\it FUSE} detector sides.
(We caution the reader than some departures over small
wavelength regions as large as ${\pm 0.02}$\,\AA~ may yet exist, 
although we have tried to correct them all.)
As a last step for conditioning {\it FUSE} data, 
we corrected the wavelengths for the most recent 
stellar radial velocity tabulated in SIMBAD.\footnote{The SIMBAD 
database is operated at the Centre Donn\'ees Astronomiques de 
Strasbourg.} The final spectra were resampled linearly to the 
original uniform mesh of the original observations, which for {\it FUSE} 
is 0.013\,\AA~ pixel$^{-1}$. The long wavelength ($>$1188\,\AA) sampling 
is 0.031\,\AA~ pixel$^{-1}$ for {\it IUE,} and 0.0052\,\AA~ pixel$^{-1}$ 
for {\it STIS}. The wavelength spacing for the far-UV region of the
{\it Copernicus} atlas is about 0.022 \,\AA~ pixel$^{-1}$. The original
wavelengths and counts from RU77 were used in our re-presentation
of their atlas.

  Next we cojoined the spectra, both below the demarcation 
wavelength 1188\,\AA~ (FUSE) and above it. 
The first step consisted of cojoining the Side 1 spectra, consisting 
of segments SiC\,1B, LiF\,1A, SiC\,1A, LiF\,2A and LiF\,1B (where
LiF\,2A was used to plug a 4\,\AA ngstrom gap in the Side 1 coverage). 
We spliced these segments at wavelengths where no conspicuous lines are
present in our template B0, B2, or B8 spectra.  The wavelengths of the
{\it FUSE} spectra are given within parentheses in Table\,2.
The spectra for each of the two {\it FUSE} sides 
are cojoined with the {\it IUE} or {\it STIS} spectra, preserving the
original uniform wavelength spacings and adopting a scale factor for 
the long wavelength spectrum that provides a continuity in flux across
the 1188\,\AA~ demarcation.\footnote{In our figure
presentations discussed below we will ``pirate" the 1181.3 to 1188\,\AA~ 
segment from Side 1 to
represent a common pair of spectra.} 

  As an aid to users of the atlas data, we denote those 
wavelengths for which ISM lines have important contributions,
particularly for the atomic and molecular hydrogen features in the
spectrum that can be contaminated by absorptions from interstellar gas. 
These lines often merge together in aggregates covering a few
\AA ngstroms. We located these wavelength regions by selecting a 
high-quality spectrum with a mixture of broad photospheric and 
contrasting sharp ISM lines. We used a coaddition of 71 {\it FUSE}
exposures of HD\,195965 (B0\,V; E$(B-V)$$\approx$ +0.41; Sasseen et al.
2002) as an ISM line template.  
We modified the ranges of the spectra for each of the 
atlas stars and forced the ISM-dominated windows to be the same between 
pairs of spectral segments covering the same wavelength ranges.

 As a last step in constructing spectra over broad wavelengths, we spliced 
{\it IUE} and {\it STIS} echelle order segments to the segment-merged 
{\it FUSE} spectra.  Splice points were selected at wavelengths where the 
local noise fluctuations of the neighboring orders were approximately equal. 
For the {\it IUE} spectral spectments we made splice points at 1188.0, 
1192.5, 1205.5, 1215.5, and 1225.0\,\AA.~ For {\it STIS} the 
splice points were more closely spaced, about every 3.5\,\AA,~ 
again starting at 1188.0\,\AA.~

\subsection{Tools for line identifications}

\subsubsection{The three template spectra}

   We selected as templates, that is representatives from which to identify 
lines over the B spectral range, the spectra of $\tau$\,Sco
(B0), HD\,37367 (B2), and HD\,182308 (B8p). 
We chose HD\,182308 rather than the cooler HD\,62714
because the lines of HD\,182308 are narrower. We used an effective
temperature of 13,100\,K 
to compute synthetic spectra for our cool-star template as this is 
close to the mean of the T$_{eff}$ values of these two cool stars 
(Table\,1).  Likewise, because the lines of
$\tau$\,Sco are among the narrowest of any nearby early B star, we chose
this as our representative B0 type. Despite the uncalibrated nature of 
its linearized fluxes, the spectral resolution and signal-to-noise ratio, 
of the {\it Copernicus} spectrum rival or exceed
the quality of the {\it FUSE} material. We determined from our
spectral synthesis model tool that the transition between the dominance 
of Fe\,III lines to Fe\,II lines occurs at about 23,000\,K.\footnote{The
transition from dominant Fe\,III
to Fe\,II lines occurs at lower temperatures in B supergiants and likely
accounts for important changes in wind characteristics (the
so-called ``bistability jump") between B0.5 and B0.7-B1 supergiants
(Crowther et al. 2006).}
This is about the expected surface
temperature of a B2\,V star.  Accordingly, we chose the HD\,37367 spectrum
as our middle-ionization template spectrum. We emphasize that the
temperatures chosen for our three spectral templates serve only 
to identify observed lines and not fit them quantitatively.

\subsubsection{Construction of the line library}

  To prepare for the identification of lines in our spectral 
templates, we compiled a line library  
from three sources: the Kurucz (1993) line library, the Vienna
Atomic Line Database (``VALD"; Piskunov et al. 1995, Kupka et al. 
1999), and the on-line atomic line database of van Hoof (2006).
The Kurucz line library has the advantage of being comprehensive and
theoretical, meaning that its coverage is not compromised in the
far-UV by absorptive optical coatings. 
However, this list has become dated. The VALD and van Hoof 
databases are periodically updated and both present a recommended
oscillator strength (log\,$gf$) value for a line 
if more than one have been published.
The VALD library is supported by an interactive web form.
The form allows the user to state
the stellar effective temperature and a line depth threshold
criterion, e.g., 1\% line depth, above which lines will be
included in a returned list. We exercised this option, and chose 
T$_{eff}$ = 23,000\,K in our requests. The van Hoof library 
is likewise extensive and is convenient to use on line, especially 
during our frequent manual cross-checking of best identifications.
Our library did not screen out lines of highly excited ions from the
Kurucz atlas contribution.
For completeness, we note that Howk et al. (2000) have suggested empirical
revisions to log\,$gf$'s of several Fe\,II resonance lines in the
range 1050\,\AA~ to 1150\,\AA~ on the basis of fittings of ISM lines in
{\it FUSE} spectra. However, the corrections they recommend are generally
within a factor of two of the log\,$gf$ values in our assembled line library.
Such corrections have little bearing on the identification of saturated 
features and could be disregarded. We were able to combine these three 
line lists and excise duplicate entries listing the same ion and nearly 
the same wavelength and atomic level excitation. The latter actions were
performed by a computer program. However, because the tolerated differences 
of wavelengths and excitations for the same line could vary by ion, some
duplications were identified and excised/ manually.  In cases of duplicate 
entries with the Kurucz list, we exercised a preference for the van Hoof 
or VALD log\,$gf$'s and wavelengths.

\subsubsection{Line synthesis }

   All our line identifications were based on our now well defined line 
library, thereby requiring all these lines have either measured or
published log\,$gf$ values in the literature. In order 
to make line identifications we used a line annotation 
facility in the spectral line synthesis program {\sc `SYNSPEC} of Hubeny, 
Lanz, and Jeffery (1994). This program can be run interactively to compute 
and plot spectral fluxes over a specified wavelength range once the user
specifies key input parameters such as metallicity (assumed to be solar), 
stellar effective temperature T$_{eff}$, log\,$g$, and microturbulence.  
In our models we used log\,$g$ = 4 and $\xi$ = 2 km\,s$^{-1}$. The T$_{eff}$
values were models closest in integral kilokelvins to the numbers in 
Table\,1.  We compared line synthesis results from standard Kurucz (1990) 
and non-LTE models calculated
by TLUSTY from the so-called ``B2006" grid (Lanz \& Hubeny 2007). The
line strengths produced from the two sets of models typically differed 
by amounts equivalent to a temperature change of 1,000\,K for models 
having a T$_{eff}$ appropriate to $\tau$\,Sco and  much smaller than 
this for a model appropriate to type B8 model.  
For the purposes of line identifications either set of models 
would serve just as well for the B8 model. In this case the low precision of 
log\,$gf$'s far outweighs uncertainties in the atmospheric T($\tau$)'s 
in the computed line strengths.

\subsubsection{Line identification methodology}

  In the last several years a few middle-UV spectral atlases have
been published with annotations for most visible spectral lines 
(e.g., Leckrone et al. 1999).  We attempted as a key part of our 
``detailed atlas" to extend this philosophy by identifying all
atomic lines responsible for far-UV absorptions in B0, B2, and B8 
main sequence spectra (and presumably for subtypes in between).
We set the short-wavelength limit of our identification list no shorter 
than at 949\,\AA,~ the starting wavelength of the RE85 line list, 
because little purpose would be served by attempting to identify 
photospheric lines to the blue of this limit, where H and H$_2$ 
lines blanketing dominates. The particular challenge faced
in the far-UV is the blending of closely spaced lines that cannot 
be resolved even in stars with rotational broadenings no larger than 
the {\it IUE} instrumental width of about 30\,km\,s$^{-1}$. 
Our procedure was to compute spectral line models in narrow 
wavelength intervals, often no larger than 3 \AA ngstroms, and 
to overplot the synthesis and the identifications provided by
{\sc SYNSPEC.} Manual intervention was often required in the
working version of the line list for the following reasons:

\begin{itemize}
\item some lines cannot be identified with certainty according to their 
published log\,$gf$ values. This is because some $gf$s are too small 
and they underpredict the line strengths, just as others are too large. 
Our procedure in such cases was first to seek the best candidate line,
next to artificially increase its log\,$gf$ by a factor of 30 in our 
working line list, and finally to rerun the {\it SYNSPEC} synthesis. 
If the line appeared in the new synthesis, 
it was marked as an uncertain identification with a symbol ``:"
in our compiled line identification lists and figures.

\item more than one line might appear as a candidate line within
our adopted resolution wavelength window:
${\pm 0.025}$\AA~ for B0 and ${\pm 0.0375}$\AA~ for B2 and B8, values
ultimately set by the instrumental resolutions and our chosen limit 
of acceptable rotational broadening (30 km\,s$^{-1}$). Typically, 
in such instances we temporarily deleted lines from our working line 
list and determined the relative strengths of the contributors, 
if necessary one by one. If they contributed by more than 50\% of
the dominant line in the blend, we retained it. We refer to
secondary lines identified within a wavelength window defining a blend 
a ``line group," and the dominant contributor is the ``primary" line 
(sometimes by a very small amount). 
Non-dominant contributors are called ``secondary" group
members. We counted lines as members if their equivalent 
widths (computed as an isolated line in {\sc SYNSPEC} and with
respect to a continuum determined by this program) was greater than the
50\% criterion just noted. In practice, almost all detectable far-UV 
lines are saturated. Thus, lines contributing less than the primary line's 
absorption do not contribute much to a line group's aggregate strength. 
We noted the relative equivalent widths of the secondary contributors, 
and our notes are available upon request.

\item a candidate line's published wavelength differed from the 
measured flux minimum value by more than $\pm{0.03}$\AA.~ In
some cases, especially for O\,II, we suspect this threshold should 
be relaxed to $\pm{0.04}$\AA. It is important to note that this limit 
does not automatically apply to two or more lines separated by more 
than the wavelength resolution bin value in nearby pairs of blended lines.  
Thus, sometimes when an array of neighboring candidate lines is included 
in our identification tables, the wavelength of the primary line 
does not closely coincide with the minimum of the blended feature.  

\end{itemize}
 
   We note that our line list also includes overpredicted
lines, that is, identifications predicted from {\sc SYNSPEC} 
with no visible counterparts in the observed spectrum. 
 
  The construction of our line list proceeded after first putting into place 
semiautomated error checking procedures. One such procedure was to check the 
ion and wavelength values against those in our line library. Our program
reported any errors in this collation, and they were corrected. The
monotonicity of the wavelengths in the list was likewise checked.
Nonetheless, we cannot claim that our list of identifications is error-free!  
The influences of some lines may yet have been over-
or underestimated. Second, we have checked our line identifications
with other published lists of prominent far-UV lines, including 
those in the Pellerin et al. (2002) atlas and the far-UV Capella 
atlas (Young et al. 2001). We noticed minor deviations in quoted 
wavelength values for several lines, and in a few cases we
could not authenticate line identifications because 
we could not find log\,$gf$ values in their secondary sources.  
The absence of any significant
discrepancies in these comparisons suggests that there are 
few or no gross or systematic errors in our list.
In all we believe our identified and unknown lines form a 
list of all essentially all the visible features that contribute 
the far-UV absorption lines of B0, B2, and B8 Galactic main 
sequence stars,
including most ISM lines.

\section{The atlas}

\subsection{The atlas and associated data products}

  This presentation of our ``detailed atlas" is different from that of
most other atlases. Typically, spectral 
atlases present a pictorial montage of all spectral 
types over the full wavelength range surveyed. Our atlas consists of 
three core products: (1) extensive line identification lists,
(2) a graphical plot of line identifications of the three template 
spectra and,
(3) data files containing merged spectra in Flexible Image Transport 
System (FITS) format.  The paper journal version of this work gives 
an abbreviated representation of the spectral plots and of the 
identification table.  
The electronic edition provides the full identification tables 
both in ASCII text and Excel ``xls" format. 
A copy of our compiled line library will also be provided upon request. 
In addition to these published venues, all products for this atlas are 
to be placed in MAST's ``High Level Science Product" (HLSP) area
(http://archive.stsci.edu/prepds/fuvbstars/), where the products are 
further vetted by MAST staff for clarity and ease of access to the 
astronomical community.

  Our detailed line lists for the three B0, B2, and B8 template 
spectra are presented in Table\,3. The table is divided into
three subpanels a-c. Each subpanel first lists the star number for which 
a line has been identified (``1," ``2," and ``3" correspond to spectral 
types B0, B2, and B8, respectively), then in two columns the identified
wavelength from our line list sources, and finally the ion identification. 
The two wavelength columns represent {\em either} the
``primary" or ``member" (secondary) of a line group, respectively, such
that one of the columns is always unfilled. 
Here ``group" refers to a common resolution window in which the wavelength 
of the primary line is located.  We found as many as 7 secondary group 
members associated with a primary line according to this definition. 
In addition to ``uncertain" identifications referred to earlier, lines
for which no extant identification and/or log\,$gf$ are unavailable
are represented in our table with the ion symbol ``UN~I" for "unknown." 
Wavelengths for
H2 identifications were taken from H2ools. The ion column also 
designates lines that may appear both in the photospheric and
ISM spectra of the template stars with the symbol ``pism."

  The rows of Table\,3 are sorted according to the primary line wavelength 
of each group and are interleaved among the three template spectra.
In the interest of portability the on-line version gives these as a
separate table for each template star.
Secondary group members follow their associated group primaries, even 
if the secondary's wavelength is slightly smaller than the primary's.
A glance at the beginning of Table\,3
suggests a trend that is indeed born out by the full listing across the
whole far-UV range: 
35 of the first short-wavelength lines on the first page are
contained in the B0 spectrum. This same page contains only 17 lines
identified for the B8 star. The reverse is also true at the longest 
wavelengths, although less markedly.

Table\,4 is a list of 167 least contaminated
lines that, typically, are visible in two or all of our template 
spectra (omitting hydrogen Lyman lines). 
The excitation of the lower level of these transitions is given in
electron Volts (eV) in the fourth column of each of the three subpanels.

Figure\,1 is a three-panel representative montage of the three template spectra 
with line identifications in the almost arbitrarily chosen region between 
1070\,\AA~ and 1077.5\,\AA. A more extended wavelength coverage would 
render the detailed line identifications difficult to read. This spectral
region is not highly contaminated by ISM H$_2$ features, as it would be
below 980\,\AA, but it also shows a small sample of them. It also exemplifies
the occasional scattered light effects that affect the edges
of spectral segments (in this case SiC2B).
Above 1148\,\AA~ 
the H$_2$ lines are no longer present in the spectra.  The region
shown in Fig.\,1  highlights the changes in line identifications 
while also including a few common lines for visual reference.  
Similarly, Figure\,2 is the first in a series of  montages of the three 
adjunct spectra, again, in the region immediately above
the 1188\,\AA~ demarcation. The top panel continues to show the {\it Copernicus}
spectrum of $\tau$\,Sco, while the second and third panels show the {\it IUE}
spectra of $\gamma$\,Peg and $\xi$\,Oct. This spectrum highlights the appearance
of the Si\,II 1193\,\AA~ and 1194\,\AA~ resonance lines, which grow in strength
from B2 through the remainder of the B spectral sequence. A series of amorphous 
C\,I lines, largely formed in the photosphere, can be seen in the lower panel. 
This is an example of the lowering of the typical ion state seen at the 
long wavelength end of the atlas.
The green line in these figures are 4-point filtered averages of the Side\,1
spectra.\footnote{We choose to plot only one side because we found it is unwise 
to average spectra for the two {\it FUSE} detector sides in an automated 
environment. Side\,1 was chosen because it provides an effective areas
equal to or surpassing the area given by Side\,2 (Kaiser \& Kruk 2009).}

Line annotations are given in Figs.\,1 and Fig.\,2 for the group primary 
lines only.  The number in parenthesis following the annotated 
identification corresponds to the {\em total} number of group members 
(whether photospheric or ISM). For example, ``(2)" means the line group  
consists of the primary and one secondary contributor.  In cases where 
the primary line is a  H$_2$ line this number is also indicated, but 
without  parentheses under the indicated red line segment.
Also, atomic ISM lines are annotated in the color green.
The electronic version of this paper contains a full set of three-panel 
spectral plots, covering our full spectral range, similar to Figs.\,1 and 2.

  The FITS products available through MAST are two-extension binary
table files. Each extension contains data for a {\it FUSE}
detector side as well as salient details about the observation
(only one extension is needed for the single {\it Copernicus} atlas
spectrum of $\tau$\,Sco). The first FITS extension contains {\it FUSE} 
Side\,1 data as well as 
data for wavelengths above 1181\,\AA.~ The spectral data are arranged 
in three columns consisting of wavelengths, fluxes, and an integer flag
with value 0, 1, or 2. 
Fluxes at wavelengths shorter than 1188\,\AA~ have a flag value 0 
if they are formed mainly in an ISM feature,\footnote{We tried
to err on the side of 
making the ISM wavelength windows large (and the photospheric ranges
correspondingly small) in order to discourage a user's identification
of features as isolated photospheric lines in any of the atlas spectra
when they may well have important ISM contributions.}, a value 1 if they
are formed mainly in the photosphere, and a value 2 (Side 1 only) if they
are associated with a wavelength greater than 1188\,\AA~ and come from 
an observation of a star other than the star oberved for wavelengths below 
this limit.  Full coverage plots can be 
made either by ignoring the flags or coplotting each set of
segments with with their separate flag codings.
The wavelength and flux vectors are in native units (wavelengths 
in \AA ngstroms, fluxes in ergs s$^{-1}$\,cm$^{-2}$\,\AA$^{-1}$).
The FITS files also 
contain header keywords that include the names, spectral types,
and E(B-V) reddening (if available) 
of the stars used, the dataset ID sequences, number of exposures, 
scaling factor to bring long wavelength ($\ge$1188 \AA ngstroms) 
fluxes into agreement with the short wavelength fluxes,
start and end times of observations, and total exposure time.

\subsection{Example: the C\,III 1176\,\AA~ region }
\label{node1176}

We exhibit as Figures 3 and 4 a montage of 8 of the 10 atlas
spectra in the region of the  C\,III 1176\,\AA~ complex. This
aggregate consists of a series of six primary C\,III lines arising
from levels at $\chi$ = 6.5\,eV. Annotations of the ions responsible
for primaries in {\em most} line
groups are given for the three templates in Fig.\,3 - at the top for B0 
and at the bottom for B2. 
These annotations are staggered 
even-odd vertically in the figure so as to make the identifications 
readable. 
In Fig.\,4 the line crowding is severe enough that 
we were obliged to display line identifications for B8 alone. 
In this case the two rows of annotations alternate at the top  
and bottom of the plot.  These figures demonstrate
that the C\,III complex forms a useful diagnostic not only for 
early B types but for late B stars as well. However, we note 
that the precise positions of individual components become
shifted by blends from nearby Fe-group line blends. 
The 1176\,\AA~ region is an especially instructive example 
of the interplay of carbon and heavy element lines.  
At least a few lines in this wavelength region 
remain visible through all B types (e.g. Cr\,III 1182\,\AA) 
or at least at the early or late ends 
of the B domain (e.g., for B0-B8: N\,I 1172 and 1177\,\AA; 
for B5-B9: C\,IV 1169\,\AA~ and Si\,III 1182\,\AA).
Some of these indicators, such as a pair of weak C\,III and C\,IV lines, 
are visible to earlier types. In fact, from a few O star 
spectra examined outside our atlas sample, we discovered 
that various C\,III lines in the wavelength interval shown in 
Fig.\,3 can form a diagnostic for differentiating between late
O spectral types and also between main sequence and giant stars 
(see $\S$\ref{chem}).

\section{Discussion}
\subsection{Line statistics}

  The number of lines found in our spectral templates for 
B0, B2, and B8p is 2288, 1612, and 2469, respectively.
Of these, 7.9\%,\footnote{This rate may be compared
to Rogerson \& Ewell's (1985) stated unidentified rate of 39\%, a
rate that is surely low because many of their identified lines still 
do have not published log\,$gf$'s.},
 2.1\%, and 2.2\% could not be identified from our line
library. 
The percentages of lines that are overpredicted 
are 12.5\%, 5.5\%, and 5.6\%, respectively.  Conversely, similar percentages,
namely 13\%, 7\%, and 6\% of our found features, have ``uncertain"
(low log\,$gf$) criterion. We expect that
the great majority of these are actually valid identifications
and their log\,$gf$s are too low. For example, random spot checks
of the log\,$gf$'s required to match the observed strengths with our
spectral syntheses suggest that the log\,$gf$ error distribution
is likewise consistent with a broad Gaussian with two approximately 
equal ``weak" and ``strong" tails. 

   Photospheric lines account for 93.7\%, 87.0\%, and 91.3\% of the
identifications in the respective templates.
The few percent complement of these are due to ISM features, which
become increasing numerous with decreasing wavelengths. For example,
shortward of 1148\,\AA, we identified a total of 200 H$_2$ 
lines that are either dominant or secondary contributors to absorption 
features. Of these only two identifications are clearly H$_2$ primary lines
in the $\tau$\,Sco spectrum, a star located along the
``Scorpius ISM tunnel." More typically, H$_2$ lines comprise the 
overwhelming contribution of the ISM component in our B2 and B8p 
template spectra and account for the lower two percentages 
(87.0\%, 91.3\%) for these cases compared to the overall identification rates. 


  As to photospheric features, we made a total of 5800 photospheric atomic
line identifications in the three template stars,
of which 4216 are unique. 
We expect that synthetic photospheric and ISM spectra could 
be computed solely from this list of unique lines.
This also means that altogether our list includes 13.7\% of our adopted 
line library tabulations over its range of 949 to 1225\,\AA. 
We believe this high rate (particularly considering the inclusion
of excited ions in our line list) speaks to the general success
of our identification efforts.  We estimate that this percentage 
would be as high as 15-16\% if we excluded wavelength regions blanketed 
by saturated resonance, hydrogen, or H$_2$ ISM lines.

  From our line statistics we find a few trends with spectral type.
The first of these is the a comparatively high rate of uncertain 
identifications and nonidentifications for B0 spectral features,
suggesting an incompleteness in log\,$gf$'s  for transitions arising 
from thrice ionized Fe-group elements. However, at least part of this 
is likely a residual of the problem that interfered with the 
identification efforts by RE85.
Many of RE85's still unconfirmed identifications may well prove 
to be correct at some future point when quantitative measurements
of these lines are made.

  A second trend with spectral type is in the number of far-UV 
photospheric lines identifed across the range of B spectra.
We found a local minimum at type $\sim$B2 in the number of detected 
lines.  Some 20-25\% of the comparative dearth of identifications
at type B2 arises from obscuration of photospheric lines by the larger 
wavelength regions dominated by strong ISM H$_2$ and photospheric resonance 
lines arising from ions like Si\,III.
The rest of this dearth is a consequence of the ionizations changing
from thrice and single ionization states to second ionizations as one moves
toward B2 from  B0 and B8. Thus, the addition of lines of twice-ionized 
states does not compensate for the loss of lines from the other two ion
stages. An additional detail is that over 91\% of the identified photospheric 
lines in the B2 spectrum are also visible (as primary or secondary
lines) in either the B0 or B8 spectrum. 
Conversely, some 67\% and 53\% of the 
lines in the B0 and B8 spectrum are {\em not} visible in the B8
and B0 spectrum, respectively. Some 608 primary or secondary lines remain
visible across the entire B main sequence domain.

  A third trend is that the percentage of isolated primary lines in 
resolution wavelength bin groups declines dramatically across the 
early B spectral range (from 81\% at B0, 66\% at B2, to 59\% at B8p) 
and also with decreasing wavelength. 
Some of these ``isolated" primary lines are barely resolved, and these 
can be utilized only in a sharp-lined spectrum. For this reason
the list of least contaminated lines (Table\,4) runs to only 5\% of the
total line list.

  The general expectation that the Fe- group lines dominate the far-UV 
spectra of hot stars is borne out by our identifications. Indeed
iron itself dominates this population.
Numbers of iron lines identified in the associated ion 
stages are exhibited in Table\,5. Lines of chromium, followed 
closely by manganese, are a distant second to the incidence of iron lines.

\subsection{Previous critique of Rogerson-Ewell identifications in $\tau$ Sco atlas}

   Cowley \& Merritt (1987; hereafter ``CM87") employed Monte 
Carlo techniques to test for the presence of several ions identified 
in the $\tau$\,Sco atlas by Rogerson \& Ewell (1985). 
CM87 have critiqued RE85's claims that lines of certain ions in 
particular are present in this spectrum. Here we offer
comments on these differences based on our own identifications
specifically in the region 949\,\AA~ to 1225\,\AA:

\no {\it N II:~ } CM87 questioned whether excited N\,II features 
are present. However, since they did not list the earlier claimed
identifications by RE85, we cannot address their objections directly.
Overall, we found 6 excited lines from this ion in our photospheric 
syntheses (including 3 resonance lines apparently not in dispute). 
Since these excited N\,II lines arise from levels at 13.5\,eV, they 
can hardly be ISM lines. Thus, the presence of photospheric 
N\,II lines in this star's far-UV spectrum is likely.

\no {\it N I:~ }  Our {\sc SYNSPEC} syntheses predicted no detectable 
N\,I photospheric resonance lines. This implies that features found at 
these wavelengths are formed in the ISM. However, the photospheric syntheses 
also predict 8 or 9 N\,I lines (one is designated uncertain). All of these 
arise from levels at 3.6\,eV. Therefore at least several photospheric 
lines from this ion are likely to be present in the far-UV $\tau$\,Sco 
spectrum as well as later types. Lines with this excitation potential
are generally not primarily formed by the ISM, but we cannot rule out 
a secondary contribution.

\no {\it Ne II:~ } RE85 noted that most of their identifications of
Ne\,II lines are blended features. 
One of these is the line at 1181.3\,\AA.~ We confirm
this, the only possible detection of this ion.  Nonetheless, 
we have designated 1181.3\,\AA~ as an uncertain identification.

\no {\it P IV:~ } CM87 reported that identifications of lines of 
this ion were insecure in this spectrum. However, we found
17 lines as P\,IV in our photospheric syntheses, all but one of 
which we consider are likely identifications. 
In addition, we confirm the detection of 
at least two P\,IV lines in the {\it IUE} wavelength range that
CM87 considered ``marginally significant." We were able to do this on
the basis of many more {\it IUE} observations made subsequent to the 
CM87 study.  The presence of this ion is secure.

\no {\it P III:~} 
CM87 questioned RE85's claim of 91 P\,III lines in this spectra. 
In the far-UV, we found only four lines of this ion.  Of these only 
1184.2 and 1194.7\,\AA~ are excited lines. Because these features are among
the strongest predicted subordinate lines for this ion in our spectral 
syntheses, we are able to confirm the identification in strong P\,III
lines in the photospheric spectrum. 

\no {\it S II:~ }  CM87 agreed with RE85 that lines of this ion are present, 
but not on the basis of as many lines they claimed. We identified 
only 4 lines for this ion, of which 2 are marked ``uncertain."  The 
presence of this ion is probably secure, but (as also for P\,III) there 
is at best a marginal hope of using the few available S\,II lines for 
diagnostic purposes.

\no {\it Mn III:~ }  CM87 stated that they found ``marginal support...
for the presence a few of the strongest Mn\,III lines." They suggested
that a future study might find the abundance of this element to be
subnormal.  However, we have identified 135 lines of this ion in the
same spectrum, of which no more than 
${\frac 13}$ are uncertain identifications. The presence of this
ion in this spectrum is certainly secure.

\no {\it Zn III:~} Like CM87 we are unable to confirm identifications
of this ion in the far-UV.


\subsection{Temperature and chemical Indicators}
\label{chem}

  In this section we indicate lines that may be used in far-UV
spectra of {\it sharp-lined} B stars near the main sequence to 
refine diagnostics of effective temperature, chemical composition, 
and occasionally log\,$g$. For the B V stars most of the
interesting abundances are likely to relate to products of the 
CNO-cycle and to evidence of Bp compositional anomalies,
usually Si, Cr, and/or Mn. However, we include lines 
of other elements that may serve as metallicity indicators.

  In our descriptions below we give weight to the ``clean" spectral 
lines listed in Table\,4. However, we also include some lines that 
have minor blend contributions if they are present over a range 
of spectral types. We made our judgments by explicitly surveying 
the B0, B1, B2, B5, B8, and B9 spectra from this atlas. 
In general, those lines of singly-ionized species exhibit 
the greatest variations in strength along the B sequence. 
Lines arising from thrice-ionized atoms, except for abundant 
elements like silicon, generally exist only in at types B0 and B1. 
Those lines exhibiting the smallest changes in equivalent width 
have moderate excitations (4-8 eV) of doubly ionized atoms 
- a delicate balance between excitation and ionization effects. 
Being least sensitive to changes in photospheric temperature, 
and not being sensitive to wind conditions, these lines are the 
best indicators of abundance. We prefix below
the ion name of Si\,IV with an asterisk because the members of the 
UV resonance doublet (1394\,\AA, 1403\,\AA) exhibit an obvious 
wind component in the great majority of early B-type spectra.  Because 
wind lines can be strongly contaminated by both emission and absorptions 
deeper in the star's atmosphere, lines listed for Si\,IV are 
especially valuable diagnostics of temperature and/or abundance.

  Reasonable ionization criteria can often be framed from the 
following ion ratios in our B star atlas spectra: carbon (IV/III/II/I), 
nitrogen (IV/III/II/I), oxygen (III/II/I), silicon (IV/III/II), 
phosphorus (V/IV/III/II), sulfur (IV/III/II), chlorine (III/II), chromium 
(IV/III/II), manganese (III/II), iron (IV/III/II), and cobalt (II/II). 
In those cases where lines of three ion stages can be detectable, all
three are generally visible only down to types B1 or B2. Exceptions 
to this statement are silicon and sulfur. For these elements lines of
three ionization states may be found in spectra for type B5 and even later.

We list lines in order of ion atomic sequence that are usually 
visible in two of our three major template spectra (B0, B2, and B8) and
which usually can be found in our Table\,3.
Lines that appear in only our B0 or B0-B1 spectra include
He\,I 958.6\,\AA, C\,IV 1107.5 and 1168.9\,\AA, P\,V 1117.9\,\AA, 
1128.0\,\AA, and Ca\,III 1116.0\,\AA. Pellerin et al. (2002) have
discussed the special case of He\,II 1084.8\,\AA~ line, whose
diagnostic value is compromised by blends of N\,II lines.
We will not repeat these in the following ion listing. 


\no {\bf C III:~ } The feature at 977.0\,\AA~ is a well known 
pressure-sensitive resonance line in B spectra, although in practice
its profile in Galactic B star spectra is marred by the appearance of
molecular H$_2$ lines (Pellerin et al. 2002, Walborn et al. 2002).  
Nonetheless, researchers can
make use of this line along with the C\,III 1176\,\AA~ complex to determine 
both an early B star's spectral type and luminosity class, particularly
in Magellanic Cloud stars well out of the Galactic plane (e.g., Walborn
et al. 2002). 
As noted in $\S$\ref{node1176}, the weak C\,III line complex between 
1165.6\,\AA~ and 1165.9\,\AA~ extends from O8 to B1-B2 on the main sequence.

\no {\bf C II:~ } Strengths of the subordinate (9.6\,eV) 961.2, 972.3, 
and 996.3\,\AA~ lines of this low-ionization species increase slowly and 
persist through B-types. Strengths of the high excitation (13.7\,eV) 
1138.9 and 1139.3\,\AA~ lines persist to B5 before weakening to invisibility.
Lines at 1009.9 and 1010.3\,\AA~ attain maximum strengths at B2-B3, 
but they then decrease slowly and are still visible at type B8.

\no {\bf C I:~} A resonance feature, 1135.8\,\AA~ is the only line clearly 
visible across all the B subtypes.  A second resonance line at 1138.5\,\AA~ 
appears at B2 and remains visible for A types and later.

\no {\bf N IV:~ } The high excitation 1131.4\,\AA~ feature is
visible from at least O9 through type B1. This appears to be a
good indicator of T$_{eff}$ above 25,000\,K. However, note that 
this line is contaminated by O\,III 1131.0\,\AA, which also fades  
at about the same rate as N\,IV. It is absent by type B2.

\no {\bf N III:~ }  The N\,III complex near 979.8\,\AA~ 
fades very fast with spectral type. As noted by Pellerin et al. (2002), 
the lines at 987.7 and 991.5\,\AA~ fade into nearby atmospheric or 
H$_2$ blends at type B2, and 1184.5\,\AA~ does so at $\sim$B5.

\no {\bf N II:~ }  The features at 1083.9\,\AA~ and in the range 
1085.5\,\AA~ to 1085.7\,\AA~ 
are resonance lines that undergo a maximum strength at about B2 and 
remain visible at B8.

\no {\bf N I:~ } The photospheric N\,I lines at 1099.0, 
1172.0, 1177.6, and 1200.2\,\AA~ have excitations in 
the range 0-3.6 eV and increase rapidly with spectral type.  The 
1184.2\,\AA~ line first makes its appearance at type B5 and strengthens
at least to early A star spectra.  Of all the N\,I lines 1200\,\AA~ is 
the cleanest in our atlas spectra.

\no {\bf O III:~ }  All five lines from this ion are highly excited 
and decrease to only marginal detectability at B2. These are at 
1016.7, 1149.6, 1157.0, 1196.7, and 1199.9\,\AA. 
Of these 1199\,\AA~ is the least blended and most suitable for 
quantitative measurement in early B spectra.
Pellerin et al. (2002) have noted the utility of 1149\,\AA~ to hotter
stars and particularly in oxygen-rich WC stars.

\no {\bf O II:~ } Only one line, the high excitation line at 1130.1\,\AA,
remains visible through the whole spectral range, attaining a maximum at
about type B8.

\no {\bf O I:~ } The line O\,I 990.2\,\AA~ is visible in all our templates.
For the $\tau$\,Sco spectrum the feature is probably a photospheric/ISM blend
Lines at 1041.6 and 1127.4\,\AA~ appear at B2 and steadily increase in strength.

\no {\bf Al II:~ } The moderate excitation line at 1048.5\,\AA~ appears 
in B2 spectra and increases in strength with advancing type.

\no {\bf $^*$Si IV:~} The most visible far-UV photospheric line 
of this important ion is 1066.6\,\AA. This feature decreases in 
strength and fades at type B6. The line at 1122.5\,\AA, as also noted by 
Pellerin et al. 2002, can be a good temperature diagnostic through B5, 
but it becomes contaminated by a strong C\,I aggregate. 

\no {\bf Si III:~ } This ion is represented by nicely contrasting 
excitation potentials (15-16 eV and 6.5 eV). The former are at 967.9,
1032.8, and 1207.5\,\AA, which decrease in strength and generally
disappear beyond type B5. Two saturated lines at 1206.5\,\AA, already 
strong at B0, become major features in late B spectra. The high excitations
of the lines at 993.5, 1109.9, and 1113.2\,\AA~ balance changes in 
silicon ionization (also noted by Pellerin et al. 2002), causing their 
strengths to undergo a maximum at B0-B2.

\no {\bf Si II:~ }  The moderate excitation (5 eV) lines 1224.2
and 1224.9\,\AA~ steadily increase in strength with type. The only 
readily visible Si\,II line in the far-UV wavelength range is 
1193.2\,\AA~ and 1194.4\,\AA, which remain visible at B2 and later types.

\no {\bf P II:~ } Two resonance lines at 1149.9 and 1152.8\,\AA~ 
steadily increase through the B types. This makes them prominent 
diagnostics in late B spectra.

\no {\bf S IV:~ } The resonance line at 1073.5\,\AA~ strengthens throughout
the B range (see also Pellerin et al. 2002).

\no {\bf S III:~ } The relatively low ionization potential of this ion
allows the resonance lines at 1201.7 and 1202.2\,\AA~ to increase
only slowly through the B types, making them useful abundance indicators.
The only S\,III line in the {\it FUSE} range, 1143.8\,\AA, is marginally 
resolved. It also increases slowly through the B range.

\no {\bf S II:~ } Several 3\,eV lines at 1095.0, 
1116.1, 1124.4, 1124.9 (strong), and 1131.6\,\AA~ make their 
appearances at B1-B2 and strengthen with advancing type. The less excited
lines at 1006.2 and 1045.7\,\AA~ increase more slowly, making them 
possible abundance indicators (along with the just noted S\,III lines).

\no {\bf Cl III:~ } The resonance line at 1015.0\,\AA~ undergoes a maximum
at type B2, beyond which it becomes contaminated by an Fe II line.
The resonance line at 1009.7\,\AA~ is visible only near type B0.

\no {\bf Cl II:~ } The line at 1087.3\,\AA, arising from a level at 3.5\,eV, is 
the only Cl\,II feature available in the far-UV, but is blended with an Fe\,II 
and Ni\,II line at B2. The far-UV Cl lines are not suitable for constraining
thermodynamic information about B-star atmospheres. Abundances of
chlorine from far-UV lines of Cl\,II or Cl\,III may be estimated only 
with great care.

\no {\bf Ti III:~ } A weak line at 1080.7\,\AA~ increases strength from
B0 to B2 but beyond B5 becomes at best semiresolved between nearby 
Fe\,II and Mn\,III lines.

\no {\bf V III:~ } A single resonance line at 1148.4\,\AA~ is visible 
throughout the B types.

\no {\bf Cr IV:~ } The high excitation (31 eV) 1103.3\,\AA~ line is 
visible in late O to B2 spectra. However, the line can be resolved 
in sharp-lined spectra.

\no {\bf Cr III:~ } The number density of Cr$^{2+}$ peaks at about B2, 
and the strengths of several available low-excitation lines increase 
noticeably. These lines include 967.5, 1040.0, 1041.3, 1047.0, 
1055.8, 1065.3, 1073.7, 1119.9, 1132.7, and 1146.3\,\AA.
Arising at 4 eV, the 1181.7 and 1187.5\,\AA~ lines slowly {\it decrease}, 
suggesting that they would make the best chromium abundance diagnostics.
This is a fortunate circumstance, e.g. for the study of abundances in
late-type chromium-rich B peculiar spectra, as these lines are just within the
wavelength limits of both {\it IUE} and {\it FUSE} high dispersion spectra.

\no {\bf Cr II:~ } There are no reliable lines in the {\it FUSE} regime 
for this ion that are visible across the entire B domain. The single 
1219.5\,\AA~ line becomes visible at about O9.5 and becomes contaminated
by blends beyond type B2.

\no {\bf Mn III:~ } The only Mn\,III line (5 eV) visible at all B 
types is 1052.7\,\AA. It increases with type at a moderate rate. 

\no {\bf Mn II:~ } Just one semi-resolved line, 1161.2\,\AA, is visible 
for this ion.  To the extent it can be resolved from nearby lines its 
strength increases slowly with type. 
Therefore it can potentially serve as an abundance
diagnostic, particularly in late B spectra where it is likely to be
used to distinguish between normal B and Mn-rich Bp populations.

\no {\bf Fe IV:~ } A few lines with excitations of 19-20 eV are 
visible in the B0 to B2 
range, and they decrease in strength with type: 1022.6, 1047.2,
1135.2, 1156.2, and 1157.4\,\AA. Of these lines 1157\,\AA~ is 
among the most useful iron ionization diagnostics in early B spectra, first 
because of its nearly constant strength and second because of its 
proximity to the strengthening Fe\,III 1157.5\,\AA~ line, which is a 
serviceable line for middle B spectra.

\no {\bf Fe III:~ } For the most part the numerous lines of Fe\,III
decrease in strength within the B range, e.g. 961.7, 997.5, 1005.1, 
1030.9, 1045.9, 1047.9, 1057.5, 1058.5, 1076.5, and 1117.3\,\AA. 
However, the features at 959.3 and 979.0\,\AA~ increase in strength slowly,
attaining a maximum at B2, and then decrease very slowly by 5-10\% at
B8.  Contrasted with other Fe\,III lines, these can be used for abundances.
Arising at 6.1 eV, the 1063.4\,\AA~ line increases in strength to B2 and
remains roughly constant at later types. Given this dependence, the line 
might be used as a secondary abundance indicator in late B stars. A few 
other lines exhibit more complicated dependences. For example 1164.7\,\AA~ 
(9.9 eV) decreases in strength from B0 to B2 and disappears at later types. 
In addition to the Fe\,IV lines noted above, this line is a useful temperature 
diagnostic for the B0-B2 stars.  Because the 1221.0\,\AA~ line undergoes a
maximum at B2, it may serve as a secondary abundance indicator in this narrow
domain and a secondary temperature indicator for early and late B spectra. 

\no {\bf Fe II:~ } All photospheric Fe II lines increase in strength 
through the B spectral domain (but note the presence of ISM lines of
both Fe\,II and Fe\,III in some B and O-type stars (Walborn et al. 2002).
Only isolated Fe\,II
lines at 994.5, 1075.6, 1115.8, 1150.6, 1160.9, and 1190.8\,\AA~ are 
visible for all B types. Several are visible only near type B2:
1091.5, 1112.0, 1116.9, 1121.9, 1124.1, 1125.4, and 1195.4\,\AA.

\no {\bf Co IV:~ }  Seven high-excitation (16-17 eV) lines are probably 
visible in late O stars and at type B0, of which three are uncertain
identifications. The likely identifications are at 1115.1, 1116.0, 
1118.2, and (to type B1) 1188.0\,\AA.

\no {\bf Co III:~ } Lines at 1049.7 and 1116.8\,\AA~ arise from low (1.9 eV) 
and high (9 eV) excitation states, respectively. 
The contrast this pair of lines offers 
to one another suggests that they could be used to form a temperature 
diagnostic. Both lines increase in strength with advancing type.

\no {\bf Ni III:~ } Because 973.7\,\AA~ increases strength through B2 to 
become blended at B5, it does not offer a reliable ionization diagnostic. 
The 979.2\,\AA~ line shows only a slight increase in strength through the B 
types. This suggests it might be used as an abundance indicator in B stars.

\no {\bf Ni II:~ }  Lines of this ion first appear at B2 and then
strengthen through at least B8. The primary lines are at 1076.0, 1134.5,
1173.4, and 1181.6\,\AA.

   It can be added that a Pt\,III line at 1080.0\,\AA~ in the HD\,182308
spectrum (B8\,Vp) is the only platinum line we have found. 
As this is the among the strongest unblended lines of this element in 
the far-UV spectrum of a B8-B9 star, 
it is not surprising to find it in a Hg-Mn B peculiar-type spectrum,
It will certainly be present in others.

Finally, the Ar\,I 1048.2\,\AA~ line is strongly contaminated by the
ISM in all or nearly all of the atlas spectra, including 
$\tau$\,Sco. In general, a number of atomic ISM lines appear in 
all three spectra, including singly ionized atoms of C, N, O, and Cl.
In spectra later than B2 it is often difficult to assess the
relative strengths of the photospheric and ISM contributions, particularly
since either contribution alone can saturate the core.

  We wish to express our appreciation to Dr. John B. Rogerson for permitting
us to include the Rogerson-Upson ({\it Copernicus}) atlas of $\tau$\,Sco in
this study. We also thank Drs. S. Adelman and C. Proffitt for providing the 
author with an unpublished, preliminary {\it IUE/SWP} atlas of the B2\,IV star
$\gamma$\,Peg. The quality of this paper was substantially improved thanks
to a number of comments by an anonymous referee. This work was supported 
by a NASA Astrophysics Data Analysis grant NNX07AH57G and a grant from 
the {\it FUSE} project NNX08AG99G to the Catholic University of America.

\clearpage

\begin{figure}[h]
  \centering
  \includegraphics[angle=180]{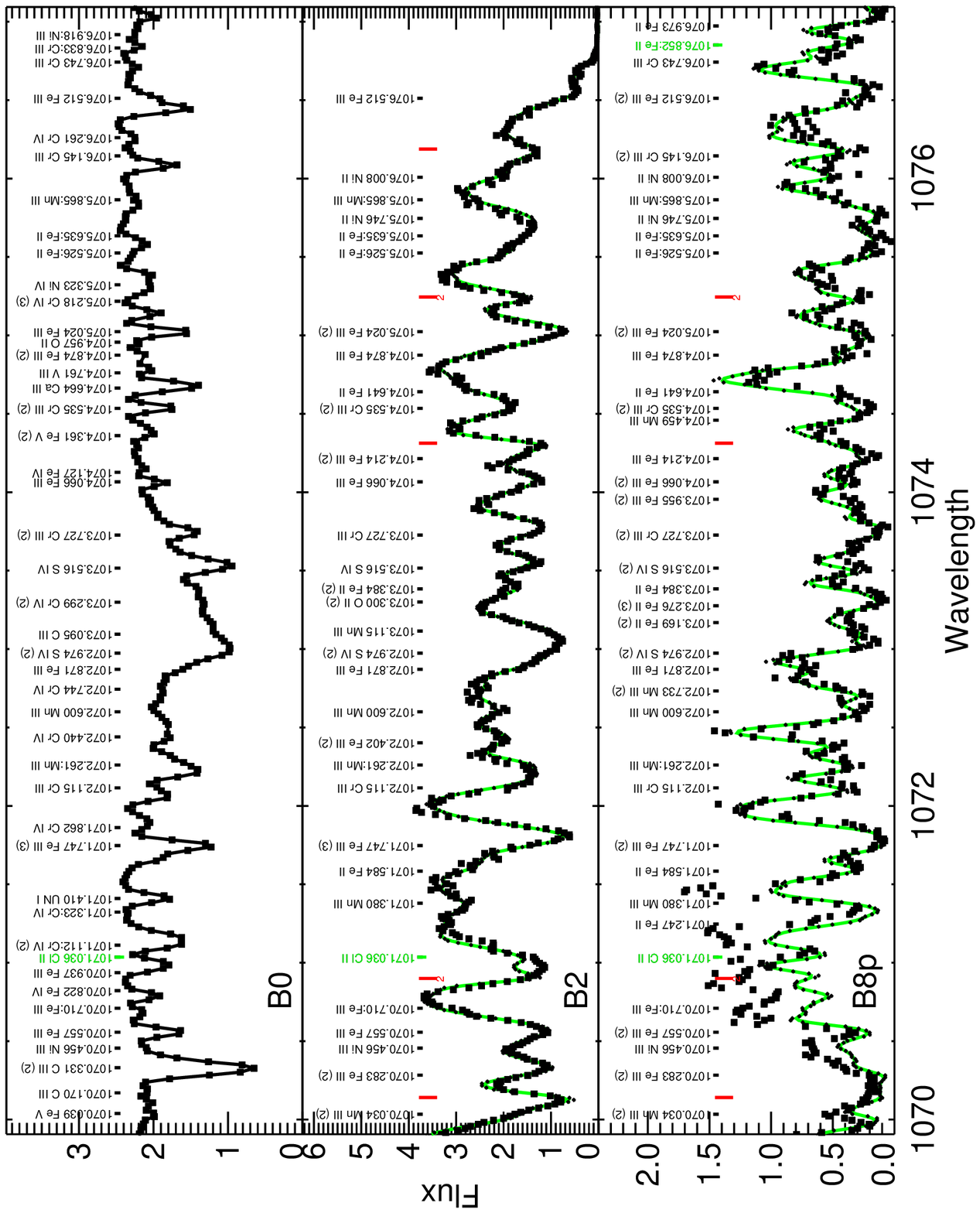}
\vspace*{0.1in}
\caption{
\scriptsize
A three panel representation of the atlas and line identifications
for main sequence B0, B2, and B8p spectra in the region 1070\,\AA~ to
1177.5\,\AA. Solid lines in the B2 and B8p panels represent a filtered 
{\it FUSE} spectrum taken from Side 1 detectors; small squares represent 
Side 2 data. 
Line identifications
are represented vertically by ion and wavelength and, where necessary,
with colons to represent uncertainties. 
Notations such as ``(3)" represent the combined number of primary 
and secondary lines in a wavelength resolution bin.
Green annotations represent atomic lines having mainly ISM 
contributions, and red vertical ticks represent ISM H$_2$ features. The
latter may also have numbers underneath them too if they are primaries in
local wavelength groups.
The ion ``UN I" represents an unidentified
absorption line. {\it Copernicus} count rates (B0: $\tau$\,Sco) 
are in units of 10$^{4}$, and {\it FUSE} fluxes in the B2 and B8p spectra 
are represented in units of 10$^{-11}$ ergs s$^{-1}$\,cm$^{-2}$\,\AA$^{-1}$.
}
\label{f1.eps}
 \end{figure}

\begin{figure}
  \centering
  \includegraphics[angle=180]{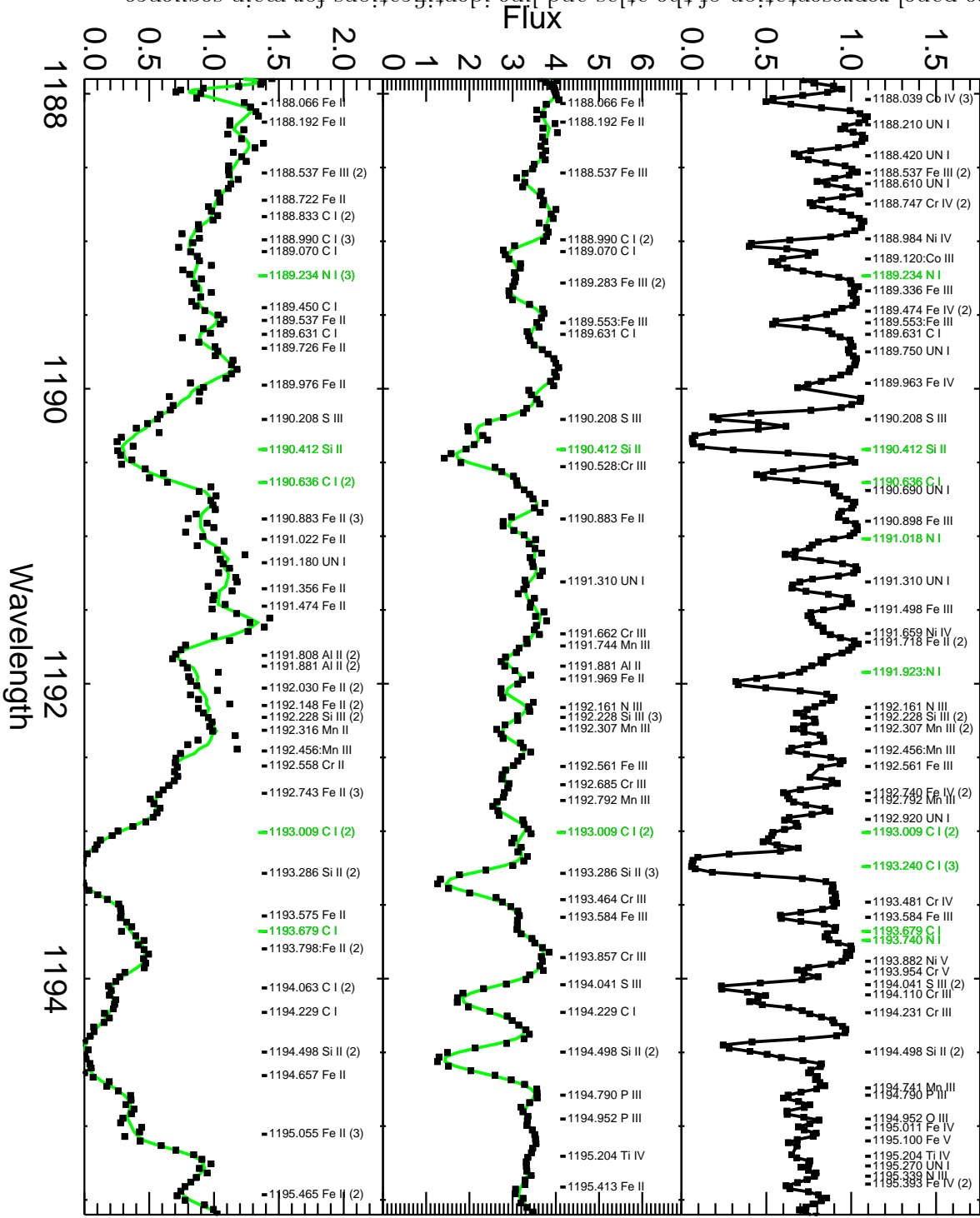}
\vspace*{-0.1in}
\caption{A three panel representation of the atlas and line identifications
for main sequence B0, B2, and B8p spectra in the region 1188\,\AA~ to
1195.5\,\AA, just above the cut-off of FUSE spectral coverage. This figure 
is a continuation of Fig.\,1 but displays spectra recorded from instruments 
other than FUSE.
}
\label{f2.eps}
 \end{figure}

\begin{figure}
  \centering
  \includegraphics[width=13cm,angle=180]{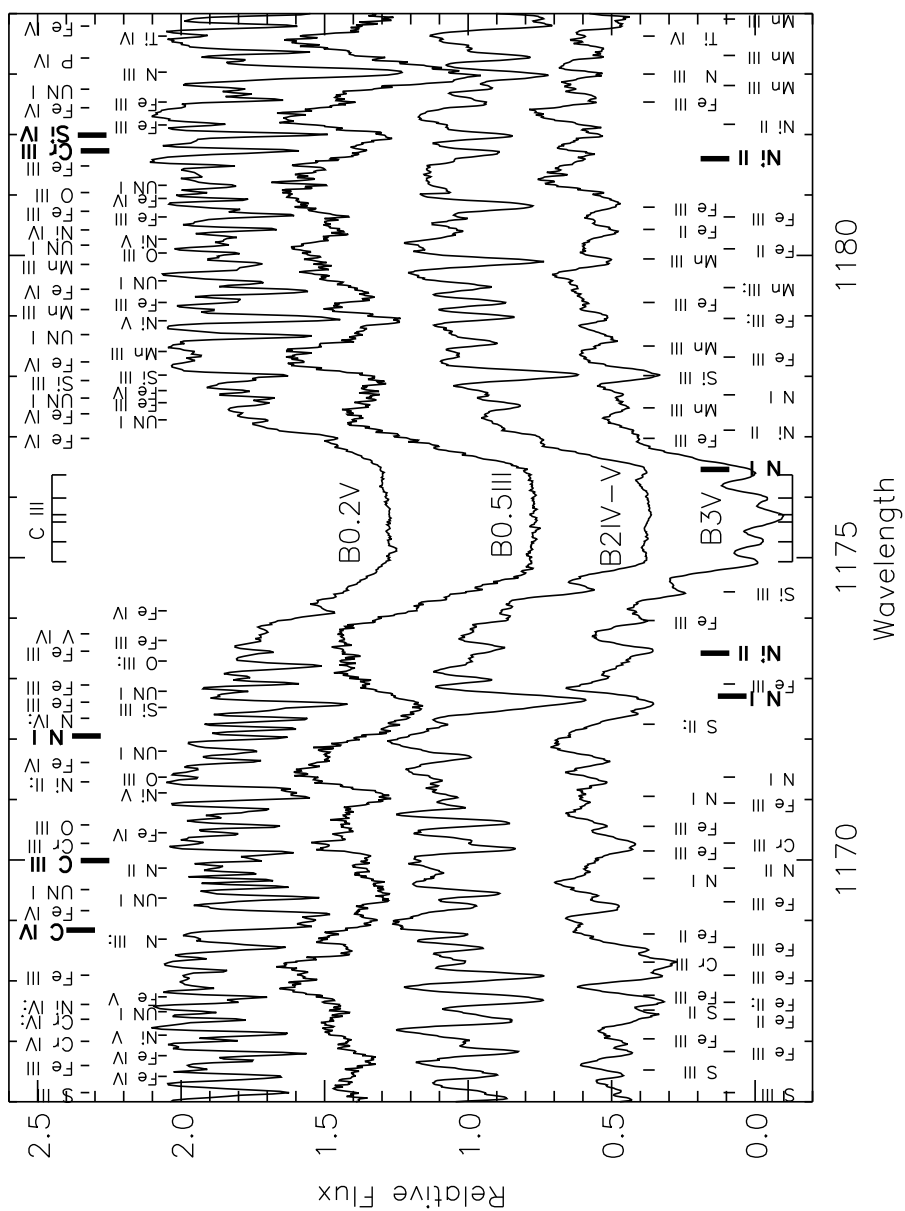}
\vspace*{-0.1in}
 \caption{A montage of the region surrounding the C\,III 1176\,\AA~
complex for the following B0-B3 main sequence stars in the atlas:
$\tau$\,Sco, HD\,102475, HD\,37367, and HD\,45057. Flux scales are c.g.s. 
and have been multiplied by 10$^{11}$, but are offset from zero for clarity.
Annotations for ``primary" identifications of {\em most} atomic line groups
are given at the top (for B0) and bottom (for B2) and are staggered for 
readability.  The C\,III and C\,IV lines indicated at the top right are, 
along with the C\,III line complex, indicators of temperature and 
gravity at B0 and into the late O stars.
The annotations at the bottom refer to the B2 spectrum, which 
is the {\it second} from the bottom in this sequence. The reader may 
discern many features that change as one proceeds from type B0 to B2. 
Other lines of Si\,III, Fe\,III, and Cr\,III lines are present throughout
the spectral type sequence, although their strengths change.
}
\label{f3.eps}
 \end{figure}

\begin{figure}
  \centering
  \includegraphics[width=13cm,angle=180]{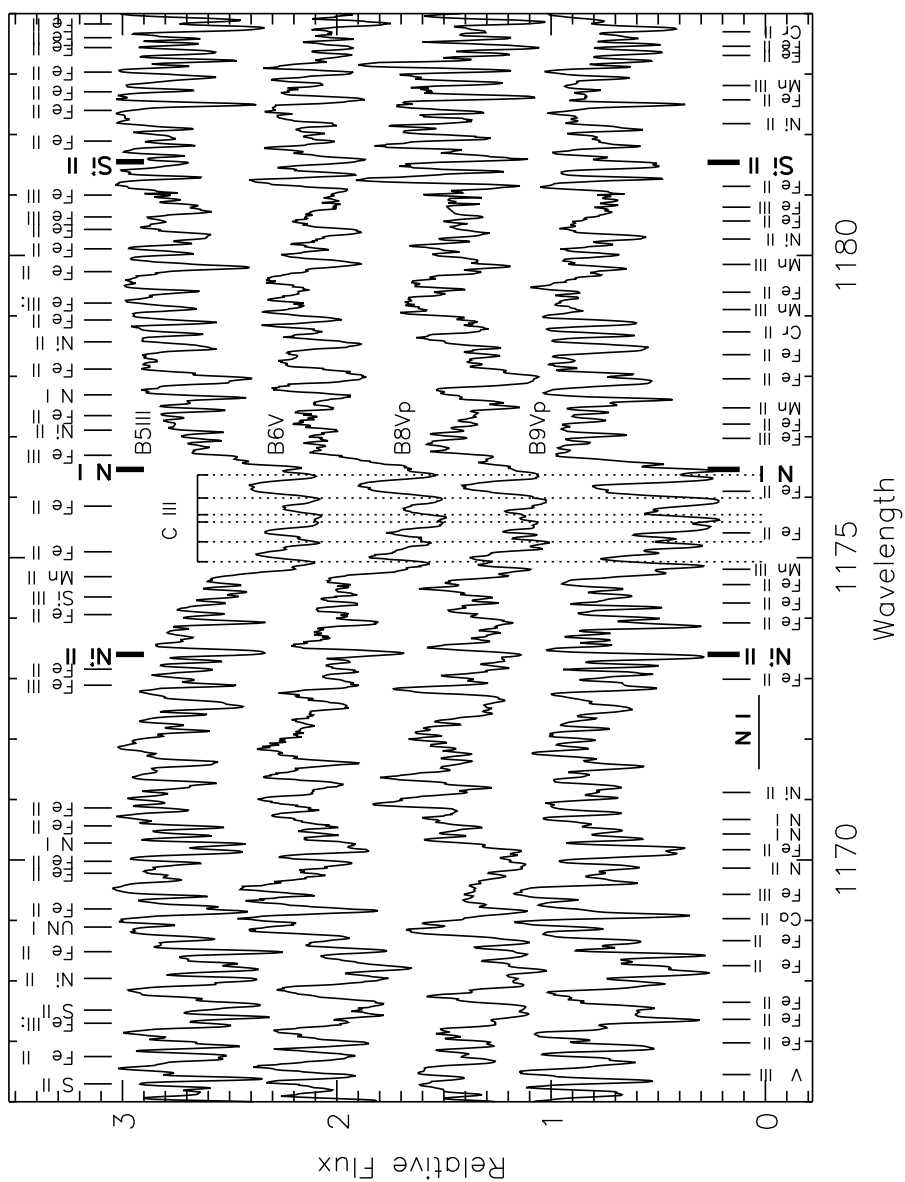}
\vspace*{-0.1in}
 \caption{A montage of the region surrounding the C\,III 1176\,\AA~
complex for four of the five B5-B9p main sequence stars in the atlas:
HD\,94144, HD\,30122, HD\,182308, and HD\,62714.  
The annotations at the top and bottom of the plot refer to the B8p
spectrum, which is the {\it third} indicated in this sequence. 
The C\,III line ``comb" is shown only at the top, with dotted lines 
tracing their positions downward through the later type spectra.
A group of N\,I lines is indicated by a single horizontal
bar. The reader should be able to discern features that change as one 
proceeds from type B5 to B9p. Other features, notably of Fe\,II and Ni\,II, 
are present throughout the sequence, although their strengths change.
}
\label{f4.eps}
 \end{figure}

\clearpage

\begin{deluxetable}{cccc|cccc}
\tablenum{1}
\tablewidth{0pt}
\tablecaption{Far UV atlas B stars }
\tablehead{

\colhead{Star} & \colhead{Sp. Type} & \colhead{E(B-V)} & \colhead{T$_{eff}$} & 
\colhead{Star} & \colhead{Sp. Type} & \colhead{E(B-V)} & \colhead{T$_{eff}$} \\
}
\startdata
$\tau$\,Sco$^*$ & B0.2\,V & 0.06 & 30,400K & HD\,201836 & B4\,IV & 0.14 & 18,620K   \\
(same$^*$)   &         &   &      & ($\iota$\,Her) & (B3\,V) &  & (17,800K) \\
HD\,113012  &  B0\,III  & 0.39 & 29,200K & HD\,94144 & B5\,III & 0.17 & 17,100K \\
($\tau$\,Sco) &       &    &     & (same)   &       &   &       \\
HD\,102475 & B0.5\,III & 0.23 & 26,100K & HD\,30122 & B6\,V & 0.23   &  15,500K \\
($\xi^{1}$\,CMa) & (B1\,III) & & (26,200K) & (same) & B6\,V &  &     \\ 
HD\,37367$^{*}$ & B2\,IV-V & 0.38 & 21,600K & HD\,182308$^*$ & B8\,Vp & -- & 13,800K  \\
($\gamma$\,Peg$^*$)&(B2\,IV)& & (22,500K)&($\xi$\,Oct$^*$) & (B8) & & (14,100K)\\
HD\,45057 & B3\,V & 0.10 & 19,100K          & HD\,62714 & B9\,Vp & 0.04 & 12,800K \\
($\zeta$\,Cas) & (B2\,IV) &  & (20,900K) & (same)   &     &  &        \\
\enddata
\tablecomments{The first line in each row corresponds to the star for
which {\it FUSE} spectra were used (except for exclusive use of the
{\it Copernicus} atlas data for $\tau$\,Sco). 
The second line in parentheses denotes cases for which 
an {\it HST/STIS} or {\it IUE} spectrum was used), often 
for a second star of identical or similar spectral type.  
Asterisks denote those stars for which spectral templates were 
adopted to identify atomic and ISM lines.}
\end{deluxetable}

\clearpage

\begin{deluxetable}{rr|rr}
\tablenum{2}
\tablewidth{0pt}
\tablecaption{Wavelength Coverage (Angstroms) of FUSE Detectors
  and Ranges Used in Cojoined Spectra }
\tablehead{

\colhead{Segment} & \colhead{Side 1} & \colhead{Segment} & \colhead{Side 2} \\ 
}
\startdata
SiC 1B & {\em 907-992} & SiC2A & {\em 917-1007}   \\
     & (930-990.45)    &       & (930-1005) \\
LiF 1A & {\em 988-1082} & LiF2B & {\em 984-1072}   \\
     & (990.45-1082)    &       & (1005-1071.35) \\
SiC 1A & {\em 1004-1092} & SiC2B & {\em 1016-1103}   \\
     & (1082-1090)    &       & (1071.35-1087.5) \\
LiF 1B & {\em 1074-1188} & LiF2A & {\em 1016-1103}   \\
     & (1094.5-1188)    &       & (1087.5-1181, \\
     &                   &       &  1090-1094.5$^*$) \\
\enddata
\tablecomments{$^*$The LiF2A segment is utilized for Side\,1 to fill 
in the gap 1090-1094.5\,\AA~ gap in that side's coverage.}
\end{deluxetable}

\clearpage

\begin{deluxetable}{rrrl|rrrl|rrrl}
\rotate
\tablenum{3}
\tabletypesize{\scriptsize}
\tablewidth{0pt}
\tablecaption{Line Identifications in B0, B2, and B8 Spectra}
\tablehead{

\colhead{(a)} & \colhead{} & \colhead{} & \colhead{} & 
\colhead{(b)} & \colhead{} & \colhead{} & \colhead{} & 
\colhead{(c)} & \colhead{} & \colhead{} & \colhead{} \\ 
\colhead{In} & \colhead{B0 prim.} & \colhead{Sec.} & \colhead{Ion} & 
\colhead{In} & \colhead{B2 prim.} & \colhead{Sec.} & \colhead{Ion} & 
\colhead{In} & \colhead{B8 prim.} & \colhead{Sec.} & \colhead{Ion} \\ 
\colhead{star \#1} & \colhead{wave.} & \colhead{wave} & \colhead{Ident} & 
\colhead{star \#2} & \colhead{wave.} & \colhead{wave} & \colhead{Ident} & 
\colhead{star \#3} & \colhead{wave.} & \colhead{wave} & \colhead{Ident} \\ 
}
\startdata
1&948.917&  &Fe\,III&  &       &  &   &   &  &   &  \\
1&949.078&  &Fe\,III&  &       &  &   &   &  &   &  \\
 &       &  &       &23&949.181&  &H2 &23 &949.181  &   &H2  \\
1&949.236&  &Mn\,III&  &       &  &   &   &  &   &  \\
1&&949.188&Fe\,III  &  &       &  &   &   &  &   &   \\
1&949.286&  &Fe\,III&  &       &  &   &   &  &   &   \\
1&&949.318&Fe\,III  &  &       &  &   &   &  &   &   \\
1&&949.328&He\,II   &  &       &  &   &   &  &   &   \\
 &&       &      &23&949.351&  &H2&23&949.351&   &H2 \\
1&949.379&  &Ar V   &  &       &  &   &   &  &   &   \\ 
1&949.544&  &Fe\,III&  &       &  &   &   &  &   &   \\
 &       &  &       &23&949.603&  &H2 &23&949.603&  H2 \\
1&949.679&  &Fe\,III&  &       &  &   &   &  &   &    \\
123&949.743&  &H I~~pism &123&949.743&  &H I~~pism&123&949.743&  &H I~~pism\\
   &       &  &          &23&&949.727&  H2            &23&    &949.727&H2\\
   &       &  &          &23&950.072&  &H2&  &       &  &       \\
   &       &  &          &  &       &  & &3&950.112&       &O I~~pism \\ 
   &       &  &          &  &       &  & &  23&       &950.072&H2 \\
   &       &  &          &23&950.314&  &H2&23&950.314&  &H2 \\ 
1&950.337&  &Fe\,III     &  &       &  &  &  &       &  &    \\
   &       &  &          &23&950.432&  &H2&23&950.432&  &H2 \\
1&950.657&  &P\,IV       &  &       &  &  &  &       &  &   \\
 &       &  &     &23&950.733&  &O I:~~ pism&23&950.733&  &O I:~~ pism\\
 &       &  &     &23&&       &H2& & & &   \\
123&950.884&  &O I~~pism&123&950.884& &     O I~~pism&123&950.884& &O I~~ pism\\
   &       &  &         &23 &       &950.816      &H2 & 23&       &950.816&H2         \\
1&950.967&  & Fe\,III &  &  &  &  &  & &  &  \\   
123&951.088&  &Fe\,III & 123 & 951.088 &  & Fe\,III & 123 & 951.088& &Fe\,III \\
1 & 951.144 &  & Mn\,III  &   &       &  &   &   &   &         &        \\
12&951.258&  &Co\,III&12&951.258&  &Co\,III  &   &       &  &        \\
1&&951.273&Fe\,III&  &  &       &  &         &   &                \\
1&951.638&  &Mn\,III &  &       &  &         &   &       &  &         \\
12&&951.619&Mn III&12&951.619&  &Mn III     &   &       &  &         \\ 
  &&       &       &23&       &951.617&H2    &23 &       &951.617&H2 \\
1&951.934&  &Mn\,III& &       &       &      &   &       &       &    \\
12&952.037&  &Co\,III&12&952.037&  &Co\,III& &  &      & \\
  &       &  &       &23&952.271&  &H2 & 23&952.271&  &H2 \\
  &       &  &       &12&  & 952.302  &Co\,III &  &   &  &   \\
12&952.302&  &Co\,III&  &  &   & &  &   &  &   \\
1&&952.304&N I~~~ISM&   &       &  &        &  &   &  &   \\
1&&952.318&O I~~~ISM&   &       &  &        &  &   &  &   \\
1&952.415&  &N I~~  pism&       &  &        &  &   &  &   \\
12&952.477&  &Fe\,III& 12&952.477&  &Fe\,III&    &    &    &  \\
1&&952.523&N\,II~~~ISM&  &   &  &  &   &    &    &         \\
12&952.729&  &Fe\,III&12&952.729&  &Fe\,III &    &    &    & \\
  &       &  &       &23&952.789&  &H2  & 23&952.789&  &H2  \\
12&952.811&  &Co\,III&12&952.811&  &Co\,III&     &    &    &  \\
12&953.002&  &Mn\,III&12&953.002&  &Mn\,III&     &    &    &   \\
1&953.134&  &Fe\,III& &   &  &   &   &    &   \\
12&953.383&  &Fe\,III&12&953.383&  &Fe\,III&     &    &    &   \\
  &       &  &       &  &       &  &       & 3&953.415&  &N I~~ pism \\
1&953.593&  &N\,IV& &   &   &    &   &   &   &  \\
\enddata
\tablecomments{Table 3 is presented in its entirety in the electronic edition
of the {\it Astrophysical Journal.}  A portion is shown here for guidance 
in data format and content.}
\end{deluxetable}

\clearpage

\begin{deluxetable}{rlr|rlr|rlr}
\tablenum{4}
\tablewidth{0pt}
\tablecaption{Unblended Lines in Far-UV B0, B2, and B8 Spectra}
\tablehead{

\colhead{B0: Wavel.} & \colhead{Ion} & \colhead{$\chi$} &
\colhead{B2: Wavel.} & \colhead{Ion} & \colhead{$\chi$ } &
\colhead{B8: Wavel.} & \colhead{Ion} & \colhead{$\chi$ } \\
 & & (eV) & & & (eV) & & & (eV) \\
}
\startdata
953.383&Fe III&2.5&953.383&Fe III&2.5&&& \\
955.334&N IV&26.8&&&&&& \\
958.698&He II&40.8&958.698&He II&40.8&&& \\
&&&958.780&Mn III&3.3&958.780&Mn III&3.3 \\
961.033&P II&0.0&961.033&P II&0.0&961.033&P II&0.0 \\
961.711&Fe III&2.7&961.711&Fe III&2.7&961.711&Fe III&2.7 \\
961.901&Fe III&4.4&961.901&Fe III&4.4&961.901&Fe III&4.4 \\
962.114&P II&0.0&962.114&P II&0.0&962.114&P II&0.0 \\
963.425&O III&2.7&963.425&O III&2.7&963.425&&2.7 \\
967.561&Cr III&2.2&967.561&Cr III&2.2&967.561&Cr III&2.2 \\
967.944&Si III&15.2&967.944&Si III&15.2&&& \\
970.024&Mn III&3.6&970.024&Mn III&3.6&970.024&Mn III&3.6 \\
971.626&Fe IV&22.7&&&&&& \\
972.365&C II&9.3&972.365&C II&9.3&977.365&C II&9.3 \\
&&&976.713&Fe III&4.4&976.713&Fe III&4.4 \\
977.020&C III&0.0&977.020&C III&0.0&977.020&C III&0.0 \\
979.032&Fe III&5.3&979.032&Fe III&5.3&&& \\
&&&&&&977.229&P II&0.1 \\
983.539&Fe III&5.3&983.539&Fe III&5.3&983.539&Fe III&5.3 \\
989.470&Fe III&0.1&989.470&Fe III&0.1&989.470&Fe III&0.1 \\
990.204&O I&0.0&990.204&O I&0.0&990.204&O I&0.0 \\
993.519&Si III&6.5&993.519&Si III&6.5&993.519&Si III&6.5 \\
994.473&Si III&6.5&994.473&Si III&6.5&994.473&Si III&6.5 \\
996.365&C II&9.3&996.365&C II&9.3&996.365&C II&9.3 \\
997.386&Si III&6.6&&&&&& \\
999.375&Fe III&0.1&999.375&Fe III&0.1&&& \\
&&&1000.084&Mn III&5.3&&& \\
1000.157&Fe IV&19.1&&&&&& \\
&&&1000.489&S II&1.8&1000.489&S II&1.8 \\
&&&1004.555&Fe III&2.7&1004.555&Fe III&2.7 \\
&&&1006.094&S II&1.8&&& \\
1007.112&Fe III&2.6&&&&1007.112&Fe III&2.6 \\
1009.986&C II&5.3&1009.986&C II&5.3&1009.986&C II&5.3 \\
1010.371&C II&5.3&1010.371&C II&5.3&1010.371&C II&5.3 \\
1012.498&S III&0.0&1012.498&S III&0.0&1012.498&S III&0.0 \\
1012.846&Mn III&4.9&&&&&& \\
1015.022&Cl III&0.0&1015.022&Cl III&0.0&&& \\
1015.554&S III&0.1&1015.554&S III&0.1&&& \\
1021.105&S III&0.1&&&&&& \\
1021.344&S III&0.1&&&&&& \\
&&&1022.072&Fe III&2.7&1022.072&Fe III&2.7 \\
1024.110&Fe III&3.8&&&&&& \\
1028.094&P IV&8.4&&&&&& \\
1028.556&Fe III&6.3&1028.556&Fe III&6.3&&& \\
1032.855&Si III&16.1&1032.855&Si III&16.1&1032.855&Si III&16.1 \\
1037.012&C II&0.0&1037.012&C II&0.0&1037.012&C II&0.0 \\
1040.050&Cr III&0.0&1040.050& Cr III&0.0&1040.050&Cr III&0.0 \\
&&&1040.687&O I&0.0&1040.687&O I&0.0 \\
1042.869&Cr III&2.2&1042.869&Cr III&2.2&1042.869&Cr III&2.2 \\
&&&1044.282&Co III&1.9&1044.282&Co III&1.9 \\
1044.755&Co III&1.8&1044.755&Co III&1.8&1044.755&Co III&1.8 \\
1049.650&P IV&27.2&&&&&& \\
&&&1051.906&Cr III&4.6&1051.906&Cr III&4.6 \\
&&&1054.312&Cr III&2.6&&& \\
1054.608&Fe IV&19.8&&&&&& \\
&&&1054.968&MnIII&5.4&1054.968&MnIII&5.4 \\
&&&1055.525&Mn III&5.4&1055.525&Mn III&5.4 \\
1057.982&Mn III&5.2&1057.982&Mn III&5.2&1057.982&Mn III&5.2 \\
1059.119&Cr III&3.2&&&&&& \\
1062.678&Si III&0.0&1062.678&Si III&0.0&&& \\
&&&&&&1064.703&Fe III&0.0 \\
1066.614&Si IV&19.9&1066.614&Si IV&19.9&1066.614&Si IV&19.9 \\
1069.496&Fe III&3.1&1069.496&Fe III&3.1&1069.496&Fe III&3.1 \\
1069.686&C III&6.5&1069.686&C III&6.5&1069.686&C III&6.5 \\
1070.331&C III&6.5&1070.331&C III&6.5&&& \\
1071.747&Fe III&3.1&1071.747&Fe III&3.1&&& \\
1073.518&S IV&0.1&1073.518&S IV&0.1&&&  \\
1073.727&C III&2.3&1073.727&C III&2.3&1073.727&C III&2.3 \\
1075.024&Fe III&3.1&1075.024&Fe III&3.1&1075.024&Fe III&3.1 \\
1076.145&Cr III&2.3&&&&&& \\
1077.143&S III&1.4&&&&&& \\
1079.384&O III&33.9&&&&&& \\
1080.779&Fe III&9.3&1080.779&Fe III&9.3&&& \\
1083.420&Fe II&0.0&1083.420&Fe II&0.0&1083.420&Fe II&0.0 \\
1084.580&N II&0.0&1084.580&N II&0.0&1084.580&N II&0.0 \\
1085.546&N II&0.0&1085.546&N II&0.0&1085.546&N II&0.0 \\
1085.701&N II&0.0&1085.701&N II&0.0&1085.701&N II&0.0 \\
&&&1086.248&Mn III&3.6&1086.248&Mn III&3.6 \\
1089.670&Fe III&6.2&1089.670&Fe III&6.2&1089.670&Fe III&6.2 \\
1090.410&C III&35.6&&&&&& \\
1093.332&Fe III&5.3&1093.332&Fe III&5.3&&& \\
&&&&&&1095.305&S II&3.0 \\
1096.606&Fe III&6.2&&&&&& \\
&&&1097.649&Fe III&6.2&&& \\
1098.917&S IV&11.7&&&&&& \\
1099.476&Cr IV&21.6&&&&&& \\
&&&&&&1100.517&Fe II&0.1 \\
1100.040&S IV&11.7&&&&&& \\
&&&1100.589&Cr III&2.6&&& \\
&&&1102.871&Cr III&2.6&&& \\
&&&1105.983&Fe III&6.2&&& \\
1106.036&N III&27.4&&&&&& \\
&&&1106.217&Fe III&6.2&&& \\
1107.591&C IV&39.7&&&&&& \\
1108.356&Si III&6.5&1108.356&Si III&6.5&&& \\
1109.940&Si III&6.5&1109.940&Si III&6.5&&& \\
1110.905&S III&25.3&&&&&& \\
&&&1111.104&Mn III&3.3&1111.104&Mn III&3.3 \\
&&&1111.212&Mn III&3.3&1111.212&Mn III&3.3 \\
1113.230&Si III&6.5&1113.230&Si III&6.5&1113.230&Si III&6.5 \\
&&&1114.549&Mn III&3.2&&& \\
&&&1117.374&Fe III&6.2&&& \\
&&&1117.891&Fe III&2.8&&& \\
1117.989&P V&0.0&&&&&& \\
1118.552&P IV&13.0&&&&&& \\
&&&1119.445&Fe III&6.2&1119.445&Fe III&6.2 \\
1121.236&Fe III&6.2&1121.236&Fe III&6.2&1121.236&Fe III&6.2 \\
&&&1124.881&Fe III&0.0&1124.881&Fe III&0.0 \\
1126.875&Fe III&0.0&1126.875&Fe III&0.0&&& \\
&&&1127.431&O I&0.0&1127.431&O I&0.0 \\
1128.340&Si IV&8.9&1128.340&Si IV&8.9&&& \\
&&&1129.191&Fe III&0.1&&& \\
1130.150&O II&12.9&1130.150&O II&14.9&1130.150&O II&14.9 \\
1130.402&Fe III&0.1&1130.402&Fe III&0.1&&& \\
&&&1132.382&O II&14.9&&& \\
1135.762&N III&30.4&&&&&& \\
1138.551&O III&26.1&&&&&& \\
1138.936&C II&13.7&1138.936&C II&13.7&1138.936&C II&13.7 \\
&&&1139.332&C II&13.7&&& \\
&&&1141.272&Fe II&7.1&&& \\
&&&1141.740&C II&7.3&&& \\
1143.874&S III&1.4&1143.874&S III&1.4&1143.874&S III&1.4 \\
1144.309&Si III&16.1&1144.309&Si III&16.1&1144.309&Si III&16.1 \\
1145.122&Si III&17.7&1145.122&Si III&17.7&&& \\
1145.669&Si III&16.1&1145.669&Si III&16.1&&& \\
&&&1146.342&Cr III&3.1&&& \\
&&&1148.591&Al II&6.4&1148.591&Al II&6.4 \\
1149.602&O III&24.4&1149.602&O III&24.4&&& \\
1149.946&P II&0.0&1149.946&P II&0.0&1149.946&P II &0.0 \\
1152.806&P II&0.0&1152.806&P II&0.0&1152.806&P II&0.0 \\
&&&1153.588&Cr III&3.2&1153.588&Cr III&3.2 \\
1153.775&O III&24.4&&&&&& \\
&&&1155.002&P II&0.0&1155.002&P II&0.0 \\
&&&&&&1155.267&Fe II&2.2 \\
&&&1155.809&Fe II&6.2&1155.809&Fe II&6.2 \\
1165.810&O III&32.2&1165.810&O III&32.2&&& \\
&&&1174.435&Si III&16.1&1174.435&Si III&16.1 \\
1174.933&C III&6.5&1174.933&C III&6.5&1174.933&C III&6.5 \\
&&&&&&1175.098&Fe III&2.8 \\
1175.263&C III&6.5&1175.263&C III&6.5&1175.263&C III&6.5 \\
1175.590&C III&6.5&1175.590&C III&6.5&1175.590&C III&6.5 \\
1175.711&C III&6.5&1175.711&C III&6.5&1175.711&C III&6.5 \\
1175.987&C III&6.5&1175.987&C III&6.5&1175.987&C III&6.5 \\
1176.369&C III&6.5&1176.369&C III&6.5&1176.369&C III&6.5 \\
1178.012&Fe III&16.1&1178.012&Fe III&16.1&1178.012&Fe III&16.1 \\
&&&1180.798&Fe III&6.1&1180.798&Fe III&6.1 \\
1181.726&Cr III&4.0&1181.726&Cr III&4.0&1181.726&Cr III&4.0 \\
1182.016&Si III&20.6&1182.016&Si III&20.6&&& \\
&&&&&&1187.957&Fe II&3.2 \\
1188.039&Co IV&16.9&&&&&& \\
&&&1188.537&Fe III&6.2&1188.537&Fe III&6.2 \\
1190.208&S III&0.0&1190.208&S III&0.0&1190.208&S III&0.0 \\
&&&1190.412&Si II&0.0&1190.412&Si II&0.0 \\
1193.283&S III&0.0&1193.283&S III&0.0&1193.283&S III&0.0 \\
&&&1194.488&Si II&0.0&1194.488&Si II&0.0 \\
1197.239&O III&36.5&&&&&& \\
1200.223&N   I&0.0&1200.223&N   I&0.0&1200.223&N   I&0.0 \\
1201.726&S III&0.1&1201.726&S III&0.1&1201.726&S III&0.1 \\
1202.212&S III&0.1&1202.212&S III&0.1&&& \\
1204.925&Cr III&2.3&1204.925&Cr III&2.3&&& \\
1206.084&Fe IV&17.2&1206.084&Fe IV&17.2&&& \\
1206.500&Si III&0.0&1206.500&Si III&0.0&1206.500&Si III&0.0 \\
1207.517&Si III&15.2&1207.517&Si III&15.2&1207.517&Si III&15.2 \\
&&&1208.451&Fe III&5.1&1208.451&Fe III&5.1 \\
1210.455&Si III&15.2&&&&&& \\
1219.192&Fe IV&17.2&&&&&& \\
1224.967&Si II&5.3&1224.967&Si II&5.3&1224.967&Si II&5.3 \\
\enddata
\end{deluxetable}

\begin{deluxetable}{lccc}
\tablenum{5}
\tablewidth{0pt}
\tablecaption{Distribution of identified far-UV Fe lines }
\tablehead{

\colhead{Ion} & \colhead{B0 } & \colhead{B2 } & \colhead{B8 }  \\

}
\startdata
Fe V &   ~43  &     &   \\
Fe IV &  502 &  ~20 &   \\
Fe III & 163 & 414 & 358  \\
Fe  II & ~29  & 117 & 707  \\
\enddata
\end{deluxetable}


\clearpage
\begin{references}
\vspace*{-.15in}

\reference{} Adelman, S. J. 1997, A\&A Suppl., 125, 65

\reference{} Adelman, S. J., Proffitt, C. R.,
et al.  2004, A\&A Suppl., 155, 179

\reference{} Blair, W. P., Oliveira, 
et al. 2009, PASP, 121, 634

\reference{} Bohlen, R. C. 1975, ApJ, 200, 402

\reference{} Cowley, A. P. 1968, PASP, 80, 453

\reference{} Cowley, C., \& Merritt, D. R. 1987, ApJ, 321, 553 (CM87)

\reference{}Croft, R. A., Hernquist, L., 
et al. 2002, ApJ, 580, 634

\reference{}Crowther, P. A., Lennon, D. J., \& Walbron, N. R. 2006,
ApJ, 446, 279

\reference{} Dawson, D., Spinrad, H., 
et al. 2002, ApJ, 570, 92

\reference{} de Mello, D. F., Leitherer, C., 
\& Heckman, T. M. 2000, ApJ, 530, 251

\reference{} de Mello, D. F., Daddi, E., 
et al. 2004, ApJ, 608, L29

\reference{} Dixon, W. V., Sahnow, D. J., 
et al. 2007, Pub. ASP, 119, 527

\reference{} Fitzpatrick, E. L., \& Massa, D. 1999, ApJ, 525, 1011

\reference{} Floquet, M. 1970, A. \& A. Suppl., 1, 1

\reference{} Garhart, M. P., Smith, M. A., Turnrose, B. E., et al. 
1997, IUE Newsletter No. 57 

\reference{} Glagolevskij, Y. V. 1994, Bull. Spec., Astrophys. Obs., 38, 152

\reference{}Hillier, D. J., \& Miller, D. L. 1998, ApJ, 496, 407



\reference{} Houk, N., \& Cowley, A. P. 1975, Michican Spectral Survey,
U. Michigan, v 1

\reference{} Howk, J. C., Sembad, K. R., 
et al. 2000, ApJ, 844, 867

\reference{}Kaiser, M. E. \& Kruk, J. 2009, ``FUSE Archival Instrument
Handbook," \penalty-10000 http://archive.stsci.edu/fuse/instrumenthandbook/

\reference{}Kilian, J. 1994, A\&A, 282, 867

\reference{} Kim Quijano, J. et al. 2007, ``STIS Instrument Handbook,"
Version 8.0, (Baltimore: STScI)

\reference{} Kupka, F., Piskunov, N.,  et al. 1999, A\&AS, 138, 119

\reference{} Kurucz, R. L. 1990, Trans. IAU, 20B, 169, http://kurucz.harvrd.edu/atoms/AEL/

\reference{} Kurucz, R. L. 1993, Kurucz CD-ROM 13 \& 22, 1994


\reference{} Lanz, T. \& Hubeny, I.. 2007, ApJS, 169, 83

\reference{} Leckrone, D. S., Proffitt, C. R., et al.
1999, AJ, 117, 1454

\reference{} Morel, T., \& Butler, K.,
et al. 2006, A\&A, 457, 651

\reference{}McCandliss, S. R. 2003, PASP, 115, 651, and tools at
http$\colon$//www.pha.jhu.edu/$\sim$stephan/h2ools2.html

\reference{}Mehlert, D., Noll, S., \& Appenzeller, I. 2002, A\&A, 393, 809

\reference{} Napiwotzki, R., Schoenberner, D., \& Wenske, V. 
1993, A. \& A., 268, 653 

\reference{} Pellerin, A., Fullerton, A. W., et al. 2002, ApJS,
143, 2002

\reference{} Pettini, M., Shapley, A. E., et al. 2001, ApJ, 554, 981

\reference{} Pettini, M., Shapley, A. E., et al. 2001, ApJ, 554, 981

\reference{} Piskunov, N., Kupka, F., et al. 1995, A\&AS, 112, 525

\reference{} Robert, C., Pellerin, C.,
et al. 2003, ApJS, 144, 21

\reference{} Rogerson, J. B., Jr., \& Ewell, M. W., Jr. 1985, ApJS, 58, 265 (RE85)

\reference{} Rogerson, J. B., Jr., \& Upson, W. L., II 1977, ApJS, 35, 37

\reference{}Rountree, J., \& Sonneborn, G., 1993,
Spectral Classification with the International Ultraviolet Explorer:
An Atlas of B-Type Spectra, NASA RP 1312

\reference{}Sassen, T. P., Hurwitz, M., 
et al. 2002, ApJ, 566, 267

\reference{}Shapley, A. E., Steidel, C. C., 
et al. 2003, ApJ, 588, 65

\reference{} Skiff, B. A. 2007, Catalog of Spectral Identifications (VizieR),
             2007yCat....102023S

\reference{} Stoehr, F. 2007, Space Telescope European Cooord. Fac. Newsl., 42, 4

\reference{} van Hoof, P. 2006 http$\colon$//physics.nist.gov/cgi-bin/AtData/main\_asd (version$\colon$ v2.04)

\reference{} Walborn, N. R. 1971, ApJS, 23, 257

\reference{} Walborn, N. R., Fullerton, A. W., et al. 2002,
ApJS, 141, 443

\reference{}Walborn, N. R., Nichols-Bohlin, J., \& Panek, R. J. 1985,
International Ultraviolet Explorer Atlas of O-Type Spectra from 1200 to 1900\,\AA,
NASA RP 1155

\reference{}Walborn, N. R., Parker, J. W., \& Nichols-Bohlin, J. 1985,
International Ultraviolet Explorer Atlas of B-Type Spectra from 1200 to 1900\,\AA,
NASA RP 1363

\reference{} Young, P. R. , Dupree, A. K., et al. 2001, ApJL, 555, L121 

\end{references}
\end{document}